# Canonical and DLPNO-Based Composite Wavefunction Methods Parametrized against Large and Chemically Diverse Training Sets. 2: Correlation-Consistent Basis Sets, Core−Valence Correlation, and F12 Alternatives

Emmanouil Semidalas and Jan M. L. Martin*

**ABSTRACT:** A hierarchy of wavefunction composite methods (cWFT), based on G4-type cWFT methods available for elements H through Rn, was recently reported by the present authors [*J. Chem. Theor. Comput.* **2020**, *16*, 4238]. We extend this hierarchy by considering the inner-shell correlation energy in the second-order Møller−Plesset correction and replacing the Weigend−Ahlrichs def2-$m$ZVPP(D) basis sets used with complete basis set extrapolation from augmented correlation-consistent core−valence triple-$\zeta$, aug-cc-pwCVTZ(-PP), and quadruple-$\zeta$, aug-cc-pwCVQZ(-PP), basis sets, thus creating cc-G4-type methods. For the large and chemically diverse GMTKN55 benchmark suite, they represent a substantial further improvement and bring WTMAD2 (weighted mean absolute deviation) down below 1 kcal/mol. Intriguingly, the lion's share of the improvement comes from better capture of valence correlation; the inclusion of core−valence correlation is almost an order of magnitude less important. These robust correlation-consistent cWFT methods approach the CCSD(T) complete basis limit with just one or a few fitted parameters. Particularly, the DLPNO variants such as cc-G4-T-DLPNO are applicable to fairly large molecules at a modest computational cost, as is (for a reduced range of elements) a different variant using MP2-F12/cc-pVTZ-F12 for the MP2 component.

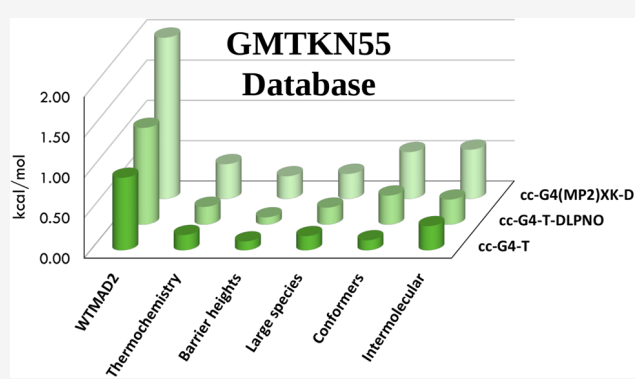

## ■ INTRODUCTION

Composite wavefunction theoretical (cWFT) methods continue to be a mainstay for reaching kcal/mol level "chemical accuracy" for reaction energies. Some of the well-established approaches include the Gaussian-n (Gn),[1−7] CBS-QB3,[8,9] multicoefficient correlation methods (MCCM),[10−12] the correlation-consistent composite approach (ccCA),[13−15] and, in sub-kcal/mol accuracy regimes, the Weizmann-n variants,[16−23] the HEAT-n methods,[24−26] and the Feller−Peterson−Dixon (FPD)[27−29] approach. All of these share a canonical coupled-cluster CCSD(T)[30,31] component. One step toward the pursuit of accurate low-cost cWFTs was a recent DLPNO-CCSD(T)-based method (DLPNO-ccCA)[32] suitable for the elements of the first and second rows of the PTE; it was parametrized to the small G2/97 training set[33,34] of 148 small closed-shell species, the largest organic molecule in it being benzene.

The above methods, in their original form, focused on light elements. Very recently, Chan, Karton, and Raghavachari (CKR)[35] extended the applicability of G4(MP2) to the entire spd blocks of H-Rn through a switch to Weigend−Ahlrichs/Karlsruhe/def2-type basis sets.[36]

When we applied this G4(MP2)-XK to the larger and more chemically diverse GMTKN55 benchmark suite—general main-group thermochemistry, kinetics, and noncovalent interactions, with 55 problem sets[37] entailing with almost 2500 unique calculations on systems as large as 81 atoms—we were astonished to find[38] WTMAD2, weighted mean absolute deviation type 2, values inferior to the best available double-hybrid[39] (see refs 40−43 for reviews) density functional theory (DFT) functionals,[43,44] which reach WTMAD2 values in the 2.2−2.3 kcal/mol range.

As it turned out, by refitting to GMTKN55 and carefully monitoring statistical significance of empirical parameters, we were able to develop[38] a new family of cWFT methods using def2 basis sets: in particular, G4-T and G4-T-DLPNO methods reached WTMAD2 values of just 1.51 and 1.66 kcal/mol, respectively. Particularly, G4-T-DLPNO is an economical option.

In these papers, inner-shell correlation was generally excluded, as the def2 basis sets are not designed for core–core (CC) and core–valence (CV) correlation. Yet it cannot be taken for granted that its neglect does not compromise accuracy. Addressing this question is the subject of the present study.

Now doing so would require switching to a different family of basis sets that have the required radial and angular flexibility in the core–valence, i.e., high-exponent, region, such as the cc-pwCVnZ (correlation-consistent, core–valence weighted, n-tuple $\zeta$ basis sets) of Peterson and Dunning.[45] But such basis sets would likely have a different convergence behavior for the valence contribution as well, and hence require reparametrization. In fact, there is evidence[21] that additional radial flexibility such as offered by core–valence basis sets benefits valence properties as well.

Basis set convergence for core–core (CC) and core–valence (CV) correlation contributions to atomization energies was recently studied in great detail in ref 46. The importance of inner-shell correlation will not be homogeneous across the GMTKN55 test suite; while it is well known (see, e.g., ref 47) that small-molecule atomization energies have inner-shell correlation components of several kcal/mol, their contribution to noncovalent interactions between first- and second-row compounds will generally be very small[48] owing to the long-distance nature of dispersion and the fairly short-range nature of CV correlation. Correlation from the $(n - 1)$d subvalence shells of Br and I is rather more important in halogen bonding.[49] For conformer equilibria and large-molecule isomerization reactions, one can expect a large degree of CC and CV cancelation between reactants and products. However, at least in principle, the inclusion of CV correlation should improve the overall GMTKN55 performance.

In this paper, we will present a hierarchy of cc-G4- and cc-G4-DLPNO-type approaches. Core–valence correlation will only be considered at the MP2 level. We will show that we can actually reduce WTMAD2 below the 1 kcal/mol threshold. Somewhat surprisingly, we find that this improvement in accuracy is due much less to core–valence correlation itself than to basis set expansion. The correlation-consistent methods deliver an attractive compromise between accuracy and computational cost for systems dominated by dynamic correlation. For systems with severe static correlation,[50] CCSD(T) is inadequate anyhow and one needs to resort to approaches such as W4,[51,52] W4-F12,[21] HEAT-III,[26,53] or FPD.[27–29]

■ COMPUTATIONAL DETAILS

All calculations were performed on the ChemFarm HPC cluster of the Faculty of Chemistry at the Weizmann Institute. Valence-only CCSD(T) calculations[30,31] were carried out using Gaussian 16, revision C.01.[54] The MP3 and RI-MP2[55,56] calculations were performed with Q-CHEM 5.2,[57] and for the DLPNO-MP2,[58] DLPNO-CCSD(T),[59] and DLPNO-CCSD(T$_1$)[60] calculations, we utilized ORCA 4.2.1.[61] Most RI-MP2-F12 calculations likewise relied on ORCA, but for technical reasons, we employed MOLPRO[62] version 2019.2 for the heavy p-block systems. The 4TB SSD (solid-state disk) arrays on the "heavyio" nodes of ChemFarm benefited the canonical MP3 and CCSD(T) calculations, even though they were insufficient for a few of the largest MP3/def2-TZVPP jobs; for those, we turned to a 40TB shared-over-InfiniBand storage server custom-developed for us by Access Technologies of Ness Ziona, Israel.

In MP2 calculations, where all electrons were correlated, we employed the following basis sets: for H and He, the correlation-consistent aug-cc-pV$m$Z[63,64] along with the corresponding RI auxiliary basis sets;[65] for the first-row (Li–Ne) and second-row (Na–Ar) atoms, the weighted core–valence basis sets aug-cc-pwCV$m$Z[45,63,64] along with the corresponding auxiliary RI ones.[66,67] The third-row alkali and alkaline earth metals K and Ca were also treated using aug-cc-pwCV$m$Z, but for molecules containing K or Ca, the RI approximation was not considered due to lack of the corresponding auxiliary basis sets. In all-electron MP2 as well as in valence-only MP2 calculations, we correlated all of the electrons of the alkaline and alkaline earth metals. For the heavy p-block elements, Ga-Kr, In-Xe, and Tl-Rn, we resorted to aug-cc-pwCV$m$Z-PP[68] (where PP stands for pseudopotential) associated with small-core multiconfigurational Dirac–Hartree–Fock (MCDHF) relativistic pseudopotentials.[69,70] As auxiliary basis sets for the heavy p-block elements, we employed cc-pV$m$Z-PP-F12/MP2FIT.[71] The cardinal number $m$ refers to T or Q, and in this text, we shall denote the core–valence basis sets used as aug-cc-pwCV$m$Z(-PP). We have also carried out a series of valence-only MP2 calculations for the species in GMTKN55 using the same aug-cc-pwCV$m$Z(-PP) basis sets. Finally, at the request of a reviewer, for comparison, we also considered the simple cc-pV$n$Z basis sets in the valence-only MP2 step, with some slight ad hoc modifications: for the anion-containing subsets AHB21, G21EA, IL16, RG18, WATER27, and BH76, we employed aug-cc-pV{T,Q}Z, and (to accommodate pseudohypervalent molecules like SO$_3$) cc-pV(n+d)Z for the second-row atoms in W4-11.[72] For technical reasons, aug-cc-pwCV{T,Q}Z was applied for MB16-43, on account of the multiple alkali and alkaline earth metals appearing in these species.

For the valence-only CCSD(T), we applied the frozen-core (FC) approximation as in the G4-type methods along with the def2-SVSP (standard def2-SVP without the polarization functions on hydrogen) and def2-TZVP basis sets of the Weigend–Ahlrichs/Karlsruhe def2 family.[36] Hence, we were able to repurpose the CCSD(T) total energies from our previous study.[38] Similarly, we repurposed MP3/def2-TZVPP along with the def2 small-core energy-consistent relativistic pseudopotentials[73] for elements heavier than Kr. The augmented def2 basis sets are available for the elements H–La and Hf–Rn; the nonaugmented ones are additionally available for Ce–Lu. We retrieved the core–valence basis sets and their auxiliary variants from the Basis Set Exchange[74] and the ccRepo basis set repository.[75] We have provided all basis sets files used in this work in the Supporting Information (SI).

The open-shell cases were treated with unrestricted HF orbitals (UHF), analogously to the reported G4-type approaches, CBS-QB3, and G4 methods. A few G4(MP2) variants use ROHF determinants,[76] which might be more appropriate for radicals prone to severe spin contamination. The dispersion model considered was that of Grimme et al.,[77] with the Becke–Johnson damping function[78] denoted as "D3(BJ)". In our previous work on double-hybrid functional parametrization,[44] we arrived at $\{a_1 = s_8 = 0, a_2 = 5.5\}$ as reasonable compromise values for the damping function's shape parameters; we retained these parameters in the present work, leaving the R$^{-6}$ overall scaling parameter $s_6$ as the single fitted parameter.

For the DLPNO-CCSD(T) and DLPNO-CCSD(T$_1$) calculations, the details are the same as in our previous work.[38] The "TightPNO" option[79] was applied throughout, and "chemical cores" were kept frozen as in ref 38, while, as with all "def2" basis sets, the deepest core electrons of elements heavier than Kr were modeled utilizing the Stuttgart−Cologne relativistic energy-consistent pseudopotentials.[73,80] The SCF convergence threshold was equal to 10$^{-9}$ au; the RIJCOSX chain-of-spheres-exchange approximation[81,82] for constructing the Fock matrices was applied, both with the default GRIDXS2 integration grid and with the most stringent GRIDX9 grid. GRIDXS2 corresponds to angular Lebedev-110 angular and radial Gauss-Chebyshev with IntAcc = 4.01, where the number of radial points is given by[83] $n_{rad} = 15 \text{IntAcc} + 5n_A - 40$, and $n_A = 30 + 5n_{rowPTE,A}$, with $n_{rowPTE,A}$ being the number of the row in the periodic table that atom A belongs to; in GRIDX9, the angular grid is Lebedev 302 and IntAcc = 4.67. We employed the def2-SVPD, def2-TZVPP, and def2-QZVPP basis sets along with the auxiliary versions of def2/J (see ref 84) and def2-SVP/C, def2-TZVPP/C, and def2-QZVPP/C (see ref 85), as stored in ORCA's internal basis set library. For the subsets AHB21, G21EA, IL16, RG18, and WATER27, we similarly applied the diffuse-function augmented def2-TZVPPD and def2-QZVPPD,[86] inspired by ref 44. For the avoidance of doubt, in the DLPNO-CCSD(T)-based or DLPNO-CCSD(T$_1$)-based cWFTs discussed here, the $E_{CCSD-MP2}$ term is calculated by subtracting the DLPNO-MP2 energy from separate single-point calculations in the same basis set, and not from the "semi-local (SL) MP2" energy reported at the post-SCF stage of a DLPNO-CCSD(T) or DLPNO-CCSD(T$_1$) run.

For the explicitly correlated RI-MP2-F12 calculations,[87] the computational details largely follow those in ref 49. We consider here the cc-pVTZ-F12 and cc-pVQZ-F12 basis sets[88] along with the corresponding auxiliary basis sets aug-cc-pVnZ/JK,[89] cc-pVnZ-F12-MP2-FIT, and cc-pVnZ-F12-OPTRI[90] (n = T or Q for cc-pVTZ-F12 or cc-pVQZ-F12, respectively), as implemented in ORCA. For molecules involving heavy p-block elements (e.g., halogen-bonded species involving bromine and iodine), the subvalence $(n − 1)d$ shell has been correlated, analogous to ref 49. For these elements, we employed the cc-pVQZ-PP-F12 basis set[71] along with the various auxiliary basis sets set from the same reference, as stored in the internal basis set library of MOLPRO 2019.2.[62] The fixed amplitude ansatz[91] is considered throughout, and the geminal exponent ($\beta$) was set equal to 1.0, as recommended in ref 92 for the RI-MP2-F12 calculations.

The primary standard for training the presented cWFT methods was the GMTKN55 benchmark;[37] as in ref 44 for the minimally empirical double hybrids, and in our previous paper on G4-T-type methods,[38] the reference data, geometries, and charge/multiplicity information were extracted from the ACCDB database of Morgante and Peverati[93] and reused verbatim (without geometry optimization). See ref 37 for details of all of the reference data, which were either CCSD(T)/CBS (i.e., extrapolation to the complete basis set limit) or higher level, many taken from previous benchmark studies in our group. These reference data are in the hypothetical motionless state without ZPE and thermal corrections. The most computationally demanding subsets C60ISO (isomerization energies of fullerene C$_{60}$ molecules)[94] and UPU23 (relative energies of uracil dinucleotides)[95] were currently not within reach for MP3/def2-TZVPP and canonical CCSD(T)/def2-TZVP. As these subsets have comparatively small weights in the WTMAD2 formula, their omission does not significantly affect WTMAD2, as has been explained at some length in ref 38. Likewise, for the standard G4, G4(MP2), and CBS-QB3 calculations carried out for comparison, these reference geometries were not reoptimized.

The calculation of the reaction energies from total energies and a reference data file, and the evaluation of WTMAD2 and associated statistics, were performed using a Fortran program developed in-house, which is available upon request. In addition, we employed the BOBYQA[96] (Bound Optimization BY Quadratic Approximation) gradient-free deterministic optimizer to optimize the energy coefficients. Numerous initial guesses were evaluated, and reoptimizations ensured that a global minimum was indeed reached.

The performance of the presented cWFT methods for the GMTKN55 database was quantified utilizing the weighted mean absolute deviation, type 2 (abbreviated WTMAD2), as defined in eq 2 of the GMTKN55 paper.[37] WTMAD2 is a function of the sizes and energy ranges of each subset

$$\text{WTMAD2} = \frac{\sum_i^{55} 56.84 \text{ kcal/mol} \cdot N_i \cdot \text{MAD}_i / |\overline{\Delta E}_i|}{\sum_i^{55} N_i} \quad (1)$$

where $N_i$ is the number of systems of subset $i$, MAD$_i$ is its mean absolute deviation from the reference values, and the average absolute reference reaction energy $|\overline{\Delta E_i}|$ ensures that errors are weighted proportionally to their importance on the varying energy scales of the different subsets.

WTMAD2 is based on MAD, a more "robust" metric in the statistical sense[97] of the word that it is less prone to perturbation by outliers, than the root-mean-square deviation (RMSD), and more suitable for the GMTKN55[37] database. Besides, WTMAD2 was the key metric in developing the double-hybrid DFT[43,44] and G4-type cWFT[38] methods.

Finally, at the request of a reviewer, we carried out a computational cost comparison with W4 theory for a small illustrative sample of molecules. For consistency, all of these calculations were carried out on identical hardware platforms, namely, two 18-core Intel Xeon Gold 6240 processors (2.30 GHz). We employed ORCA throughout the timing comparisons of cc-G4-T- and DLPNO-CCSD(T)-based methods, while for the composite method W4, we utilized MOLPRO[62] for CCSD and CCSD(T) and the MRCC[98−100] arbitrary-order coupled clyster program for the post-CCSD(T) steps.

**Description and Nomenclature of Correlation-Consistent cc-G4-Type Methods.** The standard G4-type methods share similar energy expressions with the correlation-consistent cc-G4-type ones: post-MP2 terms, particularly valence CCSD(T), are evaluated with the same def2 basis sets, in part to permit recycling the most CPU-intensive parts of the calculation from the previous work. This leaves the extrapolated Hartree−Fock reference and $E_2$ correlation energies as the key differences; both of them are calculated using the aug-cc-pwCV{T,Q}Z(-PP) basis sets. The commonly used shorthand notation cc-pV{T,Q}Z for "extrapolation from cc-pVTZ and cc-pVQZ basis sets", and analogously for aug-cc-pV{T,Q}Z, def2-{T,Q}ZVP, etc., is employed throughout the present paper.

The two-tier methods are based on the extrapolated Hartree−Fock energy, the post-HF second-order MP correction, and a CCSD(T)-MP2 component in a smaller basis set. For instance, the highly accurate G4-T method from ref 38 has the following energy expression

$$E = E_{\text{HF/def2-\{T,Q\}ZVPPD}} + [(c_{\text{E2/CBS}} + 1)$$
$$E_{\text{corr,MP2/def2-QZVPPD}} - c_{\text{E2/CBS}}E_{\text{corr,MP2/def2-TZVPPD}}]$$
$$+ c_{\text{CCSD-MP2}}(E_{\text{C,CCSD/def2-TZVP}} - E_{\text{corr,MP2/def2-TZVP}})$$
$$+ c_T E_{\text{C,(T)/def2-TZVP}} \quad (2)$$

where $E_{\text{corr,MP2}}$ is the total second-order MP correlation energy, $E_{\text{C,CCSD}}$ is the coupled-cluster single and double valence correlation energy, and $E_{\text{C,(T)}}$ is the quasi-perturbative coupled-cluster triple excitation term. The extrapolated HF energy to the basis limit in the standard G4-type methods is given in exponential form as[101]

$$E_{\text{HF/CBS}} \equiv E_{\text{HF/def2-\{T,Q\}ZVPPD}}$$
$$= \frac{E_{\text{HF/def2-QZVPPD}} - E_{\text{HF/def2-TZVPPD}}\exp(-1.63)}{1 - \exp(-1.63)} \quad (3)$$

The cc-G4 two-tier methods share similar energy expressions with their corresponding two-tier G4 variants. For the top performer, cc-G4-T, the HF/CBS using an exponential extrapolation[101] and the $E_2$ correlation energy utilizing the Schwenke-type[102,103] extrapolation, are both obtained with the correlation-consistent basis sets (aug-cc-pwCV$m$Z(-PP))

$$E = E_{\text{HF/aug-cc-pwCV\{T,Q\}Z(-PP)}} + [(c_{\text{E2/CBS}} + 1)$$
$$E_{\text{corr,MP2/aug-cc-pwCVQZ(-PP)}}$$
$$- c_{\text{E2/CBS}}E_{\text{corr,MP2/aug-cc-pwCVTZ(-PP)}}]$$
$$+ c_{\text{CCSD-MP2}}(E_{\text{C,CCSD/def2-TZVP}} - E_{\text{corr,MP2/def2-TZVP}})$$
$$+ c_T E_{\text{C,(T)/def2-TZVP}} \quad (4)$$

The extrapolated HF/aug-cc-pwCV{T,Q}Z(-PP) energy expression for all correlation-consistent cc-G4-type methods has the exponential form[101]

$$E_{\text{HF/CBS}} \equiv E_{\text{HF/aug-cc-pwCV\{T,Q\}Z(-PP)}}$$
$$= \frac{E_{\text{HF/aug-cc-pwCVQZ(-PP)}} - E_{\text{HF/aug-cc-pwCVTZ(-PP)}}\exp(-1.63)}{1 - \exp(-1.63)} \quad (5)$$

It is important to note that fitting the HF/CBS extrapolation coefficient has no perceptible effect (to two decimal places) on the WTMAD2 in cc-G4-type methods, as we found in the previous paper.[38]

We also considered the inexpensive cc-G4(MP2)-XK-type methods, which have energy expressions similar to those of CKR; however, with an HF extrapolation given by eq 5, and the scaled $E_2$ components of same-spin, $E_{\text{corr,MP2,SS}}$, and opposite-spin, $E_{\text{corr,MP2,OS}}$, obtained using the aug-cc-pwCV$m$Z-(-PP) basis sets. For instance, the energy expression of cc-G4(MP2)-XK-D becomes

$$E = E_{\text{HF/aug-cc-pwCV\{T,Q\}Z(-PP)}}$$
$$+ [c_{\text{E2,os}}E_{\text{corr,MP2,OS/aug-cc-pwCVQZ(-PP)}}$$
$$+ c_{\text{E2,ss}}E_{\text{corr,MP2,SS/aug-cc-pwCVQZ(-PP)}}]$$
$$+ (c_{\text{E(C,CCSD)}}E_{\text{C,CCSD/def2-SVSP}}$$
$$- c_{(\text{E2,os})},E_{\text{corr,MP2,OS/def2-SVSP}}$$
$$- c_{(\text{E2,ss})},E_{\text{corr,MP2,SS/def2-SVSP}}) + c_T E_{\text{C,(T)/def2-SVSP}} \quad (6)$$

We note that the coefficients of the $E_{\text{corr,MP2,OS}}$ and $E_{\text{corr,MP2,SS}}$ terms are adjustable parameters obtained together with the other parameters through minimization of WTMAD2 for the GMTKN55 database, and should not be misconstrued as identical to the original SCS-MP2.[104,105]

For the three-tier basis set methods, we similarly follow the pattern of the G4(MP3)-type variants from ref 38 but substitute HF/aug-cc-pwCV{T,Q}Z(-PP) and $E_2$/aug-cc-pwCVQZ(-PP). Both valence-only MP3 and CCSD(T) energies are retained with the same basis sets as in ref 38. Consequently, for G4(MP3)-D

$$E = E_{\text{HF/aug-cc-pwCV\{T,Q\}Z(-PP)}}$$
$$+ [c_{\text{E2,os}}E_{\text{corr,MP2,OS/aug-cc-pwCVQZ(-PP)}}$$
$$+ c_{\text{E2,ss}}E_{\text{corr,MP2,SS/aug-cc-pwCVQZ(-PP)}}]$$
$$+ c_{\text{E3}}E_{\text{[MP3-MP2]/def2-TZVPP}}$$
$$+ c_{\text{CCSD-MP3}}E_{\text{C,[CCSD-MP3]/def2-SVSP}}$$
$$+ c_T E_{\text{C,(T)/def2-SVSP}} + c_{\text{Disp}}E_{\text{[D3(BJ)]}} \quad (7)$$

The five-parameter "high-level correction" (HLC), as defined in eq 7 of ref 38 following CKR, was originally introduced in ref 3 as a correction for residual basis set incompleteness. In its original, simplest, two-parameter form introduced as part of G1 theory,[6] the two parameter values were fixed from the hydrogen atom and hydrogen molecule's exact total energies.[6] In the present work (see below), like in our previous study,[38] we found that the addition of HLC did not significantly enhance statistics, especially not at any level that would justify the introduction of five additional parameters. It indeed introduces discontinuities on bond-breaking surfaces and might otherwise jeopardize transferability to other chemical systems. Hence, none of our final recommended levels include an HLC term.

The naming of the correlation-consistent cWFTs is analogous to the original one for the standard G4-type methods (Figure 1).[38] The extrapolation of the total $E_2$ or of the individual $E_{2,\text{OS}}$ and $E_{2,\text{SS}}$, all using aug-cc-pwCV{T,Q}Z(-PP), determines the method's name. In the first case, combining the total $E_2$/CBS extrapolated in Schwenke style[102] (see ref 103 for equivalence relations with other two-point extrapolations) with CCSD(T) leads to cc-G4-n or otherwise to cc-G4-DLPNO-n if DLPNO-CCSD(T) is used instead of CCSD(T). Inserting an MP3 step for an intermediate basis set size leads to the cc-G4(MP2.X)-n variants. If we similarly scale the same- and opposite-spin $E_2$ terms, we will denote this as cc-G4-scs-n or, if an MP3 step is added, cc-G4(scsMP2.X)-n. The cardinal number n = D, T, or

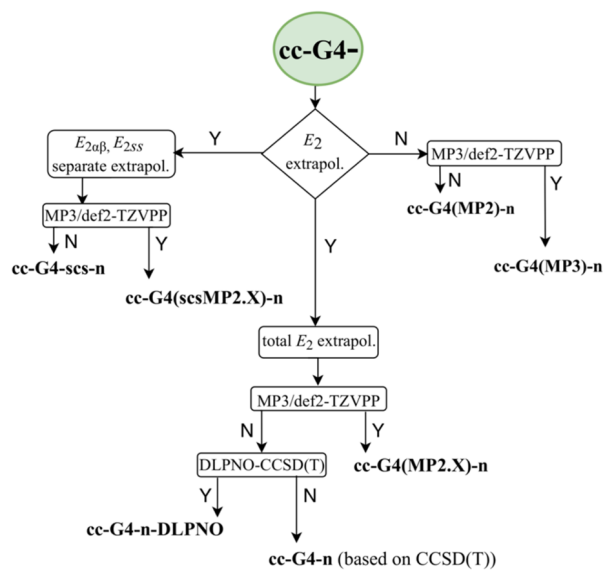

**Figure 1.** Naming scheme of the correlation-consistent cc-G4-type methods.

Q refers to the basis set employed in the CCSD(T) step, i.e., def2-SVPD, def2-TZVPP, and def2-QZVPP, respectively.

## ■ RESULTS AND DISCUSSION

The error statistics and the WTMAD2 component breakdown for selected methods (in kcal/mol) appear in Table 1. The respective abbreviations "Thermo", "Barrier", "Large", "Confor", and "Intermol" refer to the five basic subdivisions of GMTKN55: basic thermochemistry, barrier heights, reactions of large molecules, conformer equilibria, and intermolecular interactions. The table is grouped into four blocks: presently obtained "correlation-consistent" cWFT methods; cWFT from the literature; simple WFT; and the better-performing and most commonly used DFT methods.

**Effects of Basis Set Expansion and of CV Correlation Energy Inclusion.** The best approximations in Table 2 are cc-G4-T-v1 and cc-G4-T-v2, both with WTMAD2 of just 0.87 kcal/mol. cc-G4-T-v1 has four adjustable parameters, of which the fourth is the coefficient of the dispersion correction, being just −0.006. If we set it to zero instead (i.e., eliminate the dispersion correction), we obtain cc-G4-T-v2 with just three adjustable parameters (see Table S2 for the WTMAD2 contributions per subset).

This represents a substantial improvement over the WTMAD2 statistics of 1.46 and 1.49 kcal/mol, respectively, for G4-T-v1 and G4-T-v2 from ref 38. Breakdown by the five top-level subdivisions of GMTKN55 (Table 1) reveals that all five of them benefit, least so the large-molecule reactions and most so the intermolecular interactions, for which the WTMAD2 component is cut in half, from 0.63 to 0.32 kcal/mol.

Particularly for intermolecular interactions, which are a long-range phenomenon where one would intuitively expect the impact of core−valence correlation to be negligible, rationalizing this improvement entirely in terms of core−valence correlation seems implausible. But in truth, we are making two major changes at once: basis sets and core−valence correlation. Disentangling these two requires carrying out an additional set of calculations in which the same frozen-core approximation as in G4-T-v2 is applied to cc-G4-T-v2. Somewhat surprisingly, perhaps, such a cc-G4-T-v2(FC) ["frozen core"] recovers the lion's share of the improvement, at WTMAD2 = 0.94 kcal/mol.

A more detailed inspection of the individual subsets reveals that RG18 and HAL59 are the two largest contributors to the difference, with MADs reduced by 2/3. Next are BSR36 (MAD reduced by 3/4), HEAVY28 (MAD reduced by 4/10), following by a string of subsets like BH76, W4-11, G21EA, MB16-43, and the conformer subsets BUT14DIOL and AMINO20X4, for which improvements of 40−60% are seen. Because of the way subsets are weighted in WTMAD2, the small reaction energies in HAL59 and HEAVY128, and especially RG18, have an outsize contribution: the same is true to a lesser extent of BSR36 and BH76. However, while the improvement in W4-11 atomization energies does not weigh greatly in WTMAD2, the RMSD for this subset is cut in half: for small-molecule thermochemistry, an improvement from RMSD = 2.9 to 1.6 kcal/mol is significant, as is an RMSD improvement for electron affinities from 0.08 to 0.05 eV. Nevertheless, for the AMINO20X4 and BUT14DIOL conformer sets, the larger and more flexible basis sets prove very useful, although arguably, we already have quite small errors to begin with.

The difference between WTMAD2 = 0.94 kcal/mol with frozen cores, and WTMAD2 = 0.87 kcal/mol without frozen cores, implies that MP2 inner-shell correlation accounts for just 0.07 kcal/mol on WTMAD2. Half of that fairly meager improvement comes from just RSE43, where the error is cut in half. But incremental improvements for many other sets are outweighed by a deterioration in BSR36, where the valence calculation fortuitously leads to outstanding results. Still, for W4-11, RMSD drops from 1.57 to 1.27 kcal/mol, which small-molecule thermochemists would likely regard as a nontrivial improvement. It was found previously[21] in the context of the W4-F12 paper that the additional radial flexibility in core−valence basis sets is beneficial even when only valence electrons are correlated.

We considered a similar breakdown for several additional cases and consistently found that the improvement in WTMAD2 from the larger basis sets is about an order of magnitude more important than the effect of including those additional core electrons. Therefore, is their inclusion computationally wasteful? For the largest calculations in our sample, the CPU time for the largest basis set RI-MP2 calculation is, in any case, still dominated by the SCF step. The including of the additional core electrons in that step will only insignificantly add to total computational overhead; hence, we have decided to include them throughout in what follows below.

The three fitted parameters of cc-G4-T-v2 (Table 2) are $c_{MP2/CBS} = 0.671$, $c_{CCSD-MP2} = 1.054$, and $c_T = 1.126$; for its frozen-core (FC) version, $c_{MP2/CBS} = 0.787$, $c_{CCSD-MP2} = 1.051$, and $c_T = 1.139$. The pronounced change in the $c_{MP2/CBS}$ parameter (+0.116) reflects the absence of core−valence correlation energy. For the standard G4-T-v2 and 1320 systems in GMTKN55, we obtain $c_{MP2/CBS} = 0.593$, $c_{CCSD-MP2} = 1.061$, and $c_T = 1.103$; with these less complete basis sets, we find that a smaller $c_{MP2/CBS}$ coefficient (by 0.194) might compensate for the larger basis set superposition error and its effect on noncovalent interaction sets. (Note that, if the extrapolation to the CBS limit were exact, BSSE should of course vanish. For an extensive discussion of BSSE with and and G4(MP2)-XK[35] methods were obtained from ref 38. [c]D3(BJ) parameters obtained from Table S1 of ref 112. [d]$\alpha_1 = 0$, $\alpha_2 = 5.5$, $s_6 = -0.345$, $s_8 = 0$ from ref 38. [e]From ref 38, D3(BJ) parameters from Table 2 of the original D3(BJ) paper.[78] [f]The WTMAD2 component breakdown is for 1320 reactions (+63 more than in our previous work).[38]

Table 1. Statistical Errors (kcal/mol) of Recommended Methods and Selected Other WFT, cWFT, and DFT Methods for the GMTKN55 Database with the WTMAD2 Component Breakdown for the Top-Level Subsets[a]

| methods | WTMAD2 | thermo | barrier | large | confor | intermol |
|---|---|---|---|---|---|---|
| cc-G4-T-v2 | 0.87 | 0.20 | 0.09 | 0.17 | 0.11 | 0.29 |
| Ditto frozen core | 0.94 | 0.21 | 0.10 | 0.19 | 0.12 | 0.32 |
| **cc-G4-T** | 0.90 | 0.19 | 0.11 | $0.17_6$ | $0.12_5$ | 0.30 |
| Ditto frozen core | 0.99 | 0.21 | 0.12 | 0.20 | 0.13 | 0.33 |
| **cc-G4-Q-DLPNO** | 1.00 | 0.17 | 0.08 | 0.17 | 0.31 | 0.27 |
| **cc-G4-F12-T** | 1.03 | 0.20 | 0.14 | 0.24 | 0.17 | 0.28 |
| **cc-G4(MP2)-XK-T** | $1.18_5$ | 0.29 | 0.11 | 0.21 | 0.11 | 0.47 |
| **cc-G4-T-DLPNO** | $1.19_3$ | 0.22 | 0.09 | 0.21 | 0.36 | 0.31 |
| **cc-G4-F12-T-DLPNO** | $1.19_4$ | 0.25 | 0.12 | 0.22 | 0.36 | 0.24 |
| **cc-G4(MP3)-D** | 1.37 | 0.27 | 0.14 | 0.25 | 0.30 | 0.41 |
| **cc-G4-D-DLPNO** | 1.84 | 0.25 | 0.12 | 0.35 | 0.48 | 0.64 |
| **cc-G4(MP2)-XK-D** | 2.21 | 0.43 | 0.28 | 0.31 | 0.58 | 0.61 |
| cc-MP2.X-Q | 2.89 | 0.57 | 0.70 | 0.62 | 0.60 | 0.40 |
| cc-MP2.X-T | 3.09 | 0.60 | 0.64 | 0.63 | 0.61 | 0.61 |
| G4(MP2)-XK-T[38,f] | 1.42 | 0.39 | 0.16 | 0.18 | 0.13 | 0.56 |
| G4-T-v1[38,f] | 1.46 | 0.31 | 0.16 | 0.22 | 0.16 | 0.61 |
| G4-T-v2[38,f] | 1.49 | 0.32 | 0.15 | 0.23 | 0.17 | 0.63 |
| G4-Q-DLPNO[38] | 1.52 | 0.25 | 0.12 | 0.20 | 0.46 | 0.49 |
| G4-T-DLPNO[38] | 1.66 | 0.26 | 0.12 | 0.24 | 0.52 | 0.52 |
| G4(MP3)-D[38] | 1.65 | 0.37 | 0.17 | 0.28 | 0.30 | 0.55 |
| G4(MP3|KS)-D[38] | 1.96 | 0.41 | 0.28 | 0.26 | 0.45 | 0.56 |
| G4(MP2)-XK-D[38] | 2.56 | 0.46 | 0.29 | 0.34 | 0.68 | 0.79 |
| G4[3,b] | 2.52 | 0.38 | 0.23 | 0.75 | 0.38 | 0.78 |
| G4(MP2)[4,b] | 2.96 | 0.53 | 0.34 | 0.91 | 0.33 | 0.85 |
| CBS-QB3[8,9,b] | 3.10 | 0.40 | 0.35 | 0.60 | 0.20 | 1.55 |
| MP2.X-Q[38] | 3.29 | 0.71 | 0.78 | 0.88 | 0.42 | 0.50 |
| rev-G4MP2XK[38] | 3.53 | 0.50 | 0.29 | 0.61 | 1.16 | 0.96 |
| G4(MP2)-XK[35,b] | 3.71 | 0.45 | 0.31 | 0.67 | 1.25 | 1.02 |
| MP2.X-T[38] | 3.78 | 0.76 | 0.81 | 0.89 | 0.51 | 0.81 |
| SCS-MP2-D3[105,c] | 5.22 | 1.23 | 0.95 | 1.39 | 0.91 | 0.75 |
| SCS-MP2[105] | 5.35 | 0.94 | 1.01 | 1.15 | 1.02 | 1.23 |
| MP2-D3[c] | 5.83 | 1.21 | 1.21 | 1.66 | 0.87 | 0.87 |
| MP2-D3[d] | 5.54 | 1.20 | 1.18 | 1.52 | 0.80 | 0.84 |
| MP2 | 6.91 | 1.21 | 1.23 | 1.78 | 1.47 | 1.21 |
| HF-D3[e] | 13.08 | 5.05 | 2.65 | 2.06 | 1.85 | 1.48 |
| HF | 29.46 | 5.87 | 3.74 | 3.66 | 7.27 | 8.92 |
| ωB97M(2)[106] | 2.19 | 0.44 | 0.26 | 0.42 | 0.58 | 0.49 |
| xrevDSD-PBEP86-D4[44] | 2.26 | 0.56 | 0.27 | 0.52 | 0.43 | 0.47 |
| revDSD-PBEP86-D4[44] | 2.33 | 0.56 | 0.31 | 0.58 | 0.41 | 0.48 |
| revDOD-PBEP86-D4[44] | 2.36 | 0.59 | 0.30 | 0.59 | 0.41 | 0.47 |
| revDSD-PBEP86-NL | 2.44 | 0.55 | 0.30 | 0.55 | 0.47 | 0.57 |
| revDSD-PBE-D4[44] | 2.46 | 0.65 | 0.35 | 0.53 | 0.43 | 0.50 |
| revDSD-PBEP86-D3[44] | 2.42 | 0.54 | 0.31 | 0.55 | 0.46 | 0.57 |
| revDSD-BLYP-D4[44] | 2.59 | 0.57 | 0.34 | 0.58 | 0.48 | 0.62 |
| DSD-SCAN-D4[44] | 2.64 | 0.60 | 0.40 | 0.62 | 0.45 | 0.56 |
| DSD-PBE-D4[107] | 2.64 | 0.61 | 0.39 | 0.56 | 0.53 | 0.54 |
| DSD-PBEP86-D4[107] | 2.65 | 0.54 | 0.37 | 0.63 | 0.55 | 0.56 |
| revDSD-PBEB95-D4[44] | 2.70 | 0.64 | 0.31 | 0.45 | 0.78 | 0.52 |
| DSD-BLYP-D4[44] | 2.83 | 0.58 | 0.38 | 0.59 | 0.68 | 0.60 |
| DSD-PBEP86-D3[107] | 3.10 | 0.55 | 0.45 | 0.49 | 0.65 | 0.97 |
| DSD-PBE-D3[107] | 3.17 | 0.66 | 0.41 | 0.54 | 0.73 | 0.83 |
| B2GP-PLYP-D3[108] | 3.19 | 0.63 | 0.42 | 0.66 | 0.64 | 0.85 |
| ωB97M-V[109] | 3.29 | 0.73 | 0.45 | 0.64 | 0.90 | 0.57 |
| ωB97X-V[110] | 3.96 | 1.02 | 0.56 | 1.07 | 0.73 | 0.58 |
| M06-2X-D3(0)[111] | 4.79 | 0.86 | 0.48 | 1.08 | 1.22 | 1.14 |
| B3LYP-D3 | 6.50 | 1.31 | 1.14 | 1.66 | 1.15 | 1.24 |

[a]D3(BJ) is abbreviated as D3 in this table; M06-2X was evaluated with a D3(0) correction, to make D3BJ parameters and the WTMAD2 value of M06-2X without D3(0) identical. Tabulated data for the DFT methods employing the def2-QZVPP basis set (def2-QZVPPD for subsets AHB21, G21EA, IL16, RG18, and WATER27) were obtained from refs 43, 44, while the WFT (MP2, SCS-MP2, and HF) data in the same basis sets were obtained from ref 38 as were all cWFT results without inner-shell correlation. [b]The results from the conventional G4,[3] G4(MP2),[4] CBS-QB3,[8,9]

**Table 1. continued**

without counterpoise corrections, in noncovalent test sets similar to those in GMTKN55, see refs 113−115).

Next, setting $c_{CCSD-MP2} = c_T$ slightly increases WTMAD2 by 0.03 kcal/mol but leaves us with just two parameters (cc-G4-T-v6), where $c_{MP2/CBS} = 0.626$, $c_{CCSD-MP2} = c_T = 1.029$. The standard G4-T-v6 had WTMAD2 = 1.52 kcal/mol and $c_{MP2/CBS} = 0.557$, $c_{CCSD-MP2} = c_T = 1.044$. Therefore, since the statistics are not substantially affected, it is beneficial to treat all post-MP2 corrections as a single correction in cc-G4-T-v6. Figure 2 depicts the contribution to the WTMAD2 (kcal/mol) of each subset for the best two-tier methods.

We might indeed go one step further and set $c_{CCSD-MP2} = c_T = 1.0$, leaving just a single adjustable parameter. The resulting method is arguably akin to the ccCA approach without relativistic corrections. With $c_{MP2/CBS} = 0.642$, this "inspired by ccCA" composite approach with inner-shell correlation, cc-G4-T-v7 has WTMAD2 = 0.922 kcal/mol. Its counterpart cc-G4(FC)-T-v7 without core−valence correlation has WTMAD2 = 0.993 kcal/mol for $c_{MP2/CBS} = 0.727$. However, we could take this one final step and replace the one-parameter two-point extrapolation with the parameter-free three-point extrapolation combo in ccCA-PS3.[14] For such a "quasi-ccCA", we obtain WTMAD2 = 1.01 kcal/mol without inner-shell correlation, and for quasi-ccCA(noFC) without frozen cores, WTMAD2 = 0.92 kcal/mol. It is quite intriguing from a scientific point of view that in the guise of cc-G4-T-v2, we obtained something not dissimilar from ccCA from a completely different angle and the comparatively small improvement in WTMAD2 obtainable by introducing the adjustable parameters represents a "feather in the cap" of the original designers of ccCA.

At this stage, we will examine whether, for the frozen-core MP2, the larger correlation-consistent aug-cc-pwCVnZ(-PP) basis sets have a significant edge over the smaller non-augmented cc-pVnZ basis sets. We found WTMAD2 = 0.993 kcal/mol in cc-G4(FC)-T-v7 and $c_{MP2/CBS} = 0.728$ as the single adjustable parameter for MP2/aug-cc-pwCV{T,Q}Z(-PP) (note that $c_{CCSD-MP2} = c_T = 1.0$). When reducing the MP2 basis sets to cc-pV{T,Q}Z while retaining all post-MP2 components of cc-G4(FC)-T-v7, WTMAD2 more than doubles to 2.007 kcal/mol. The most notable improvements when using larger basis sets lie in the HAL59, HEAVY28, RG18, and W4-11 subtests, and to a lesser extent in S66. Now for noncovalent interaction sets, inner-valence flexibility (i.e., the pwC component of aug-cc-pwCVnZ) will not be very beneficial since we are dealing with long-range interactions. It is very well known, however (see, e.g., ref 115 and references therein), that diffuse functions, i.e., the "aug-" component, significantly reduces basis set superposition error provided the underlying basis set is not too small. In W4-11, on the other hand, we are dealing with short-range covalent bonds, and it was previously shown in the W4-F12 paper[21] that the valence correlation contribution to total atomization energies benefits from additional radial flexibility in the basis sets. All individual contributions to the WTMAD2 per subset are listed in Table S10 along with the relative energies per reaction for the two G4-T-type methods with aug-cc-pwCVnZ or cc-pVnZ basis sets in MP2(FC).

Do the energy expressions of the G4-T variants, which are less empirical than G4(MP2)-XK-type variants, hold any material advantages over the G4(MP2)-XK-T ones when including the inner-shell energy? The WTMAD2 of standard G4(MP2)-XK-T (six parameters) is 0.07 kcal/mol lower than that of G4-T (three parameters). The additional parameters of G4(MP2)-XK-T are due to the separate scaling of $E_{2,OS}$ and $E_{2,SS}$ terms with large and small basis sets. In the correlation-consistent methods, cc-G4-T-v2 comes with WTMAD2 = 0.87 kcal/mol and three parameters, and it surpasses cc-G4(MP2)-XK-T-v2 by 0.32 kcal/mol. The $E_2/\{T,Q\}$ extrapolation and the incorporation of the triples term in cc-G4-T variants are clearly adequate to recover a significant part of electron correlation instead of separately scaling the $E_{2,OS}$ and $E_{2,SS}$ components.

When reducing the CCSD(T) basis set to the smaller def2-SVSP in the two-tier methods, cc-G4(MP2)-XK-D-v1 yields the best result with WTMAD2 = 2.21 kcal/mol and seven parameters. Eliminating the dispersion term raises WTMAD2 by 0.25 kcal/mol. Switching from def2 to cc basis sets and including the CV correlation energy together improve the WTMAD2 by 0.35 kcal/mol (cc-G4(MP2)-XK-D-v1 vs G4(MP2)-XK-D-v1) and 0.27 kcal/mol (cc-G4(MP2)-XK-D-v2 vs G4(MP2)-XK-D-v2), with and without dispersion, respectively. The subsets that present the greatest improvement from G4(MP2)-XK-D-v1 to cc-G4(MP2)-XK-D-v1 are AMINO20X4, BHPERI, HAL59, MB16-43, PCONF21, S66, and TAUT15 (see Table S3 in the Supporting Information).

By scaling the $E_{2,SS}$, $E_{2,OS}$, and $E_{CCSD-MP2}$ terms, we can eliminate two semiredundant parameters at the expense of $\Delta$WTMAD2 = 0.14 kcal/mol, attaining 2.35 kcal/mol for cc-G4(MP2)-D-v1. The switch from def2 to cc basis sets, and the incorporation of the CV correlation energy, together lower WTMAD2 by 0.33 kcal/mol for cc-G4(MP2)-D-v1 relative to G4(MP2)-D-v1 (see Table S4 in the Supporting Information), the same subsets as above being most affected.

For the top-performing methods, deviations from reference reaction energies for individual data points are available in the Supporting Information. The most considerable deviations occur in the MB16-43 subset, where MB stands for the self-described "mindless benchmarking" dataset of Korth and Grimme,[116] i.e., machine-generated artificial structures (a handful of which suffer from severe spin contamination). Their reference energetics had been reevaluated in the GMTKN55 article,[37] at the W1-F12 level of theory.

**Including the CV Correlation Energy in the Three-Tier Methods.** Reducing the basis set of CCSD(T) from def2-TZVPP to split-valence greatly reduces the overall computational cost, and when a scaled MP3/def2-TZVPP correction was added as a middle tier, G4(MP3)-D was obtained, with WTMAD2 = 1.65 kcal/mol and six parameters.[38] Said three-tier method employed the post-HF components from MP2/def2-QZVPPD, a scaled MP3−MP2 difference with the def2-TZVPP basis set, and the $E_{CCSD(T)-CCSD}$ and $E_{CCSD-MP3}$ differences from CCSD(T) using the smallest basis set def2-SVSP (i.e., def2-SVP without p polarization functions on hydrogen atoms). The present three-tier methods substitute HF and MP2 with aug-cc-pwCVQZ(-PP) basis sets and retain the other components, with energy expressions of the correlation-consistent methods following the previously reported three-tier approaches (aside from the inner-shell correlation energy being included in the MP2 part).

Table 2. WTMAD2 (kcal/mol) and Optimized Parameters of Selected Standard and Correlation-Consistent cWFT Methods[d,e,f,g,h]

| MP2 | MP3 | CCSD(T) | $N_{param}$ | method | WTMAD2 | energy components coefficients | | | | | | | | HLC |
|---|---|---|---|---|---|---|---|---|---|---|---|---|---|---|
| | | | | | | $c_{(E2,os)}$ | $c_{(E2,ss)}$ | $c_{E2,os}$ | $c_{E2,ss}$ | $c_{E(C,CCSD)}$ | $c_T$ | $c_{Disp}$ | | |
| T | | D | 12 | G4(MP2)-XK[b] | 3.71 | 1.131 | 0.512 | 1.041 | 0.704 | 1.048 | 0.526 | [0] | Y |
| T | | D | 12 | revG4(MP2)-XK-H6-v1 | 3.53 | 1.307 | 0.385 | 1.170 | 0.614 | 0.984 | 0.736 | [0] | Y |
| Q | | D | 7 | G4(MP2)-XK-D-v1[b] | 2.56 | 1.309 | 0.674 | 1.124 | 0.890 | 1.113 | 0.699 | −0.383 | |
| Q | | D | 6 | G4(MP2)-XK-D-v2[b] | 2.73 | 1.482 | 0.271 | 1.320 | 0.457 | 1.052 | 0.778 | [0] | |
| Q[a] | | D | 7 | cc-G4(MP2)-XK-D-v1 | 2.21 | 1.379 | 0.414 | 1.187 | 0.617 | 1.094 | 0.854 | −0.358 | |
| Q[a] | | D | 7 | Ditto (with MP2(FC)) | 2.25 | 1.307 | 0.510 | 1.118 | 0.751 | 1.103 | 0.726 | −0.418 | |
| Q[a] | | D | 6 | cc-G4(MP2)-XK-D-v2 | 2.46 | 1.392 | 0.220 | 1.239 | 0.384 | 1.061 | 0.713 | [0] | |
| Q[a] | | D | 6 | Ditto (with MP2(FC)) | 2.52 | 1.441 | 0.147 | 1.284 | 0.327 | 1.050 | 0.756 | [0] | |
| Q | | T | 6 | G4(MP2)-XK-T-v2[c] | 1.42 | 1.624 | 0.811 | 1.526 | 0.833 | 1.070 | 0.959 | [0] | |
| Q | | T | 6 | cc-G4(MP2)-XK-T-v2 | 1.19 | 1.317 | 0.586 | 1.234 | 0.597 | 1.059 | 0.976 | [0] | |
| Q | | T | 6 | Ditto (with MP2(FC)) | 1.24 | 1.442 | 0.438 | 1.345 | 0.453 | 1.059 | 0.995 | [0] | |

| | | | | | | $c_{E2,ss}$ | $c_{(E2/CBS)}$ | $c_{E2,os}$ | $c_{CCSD-MP2}$ | | $c_T$ | $c_{Disp}$ | |
| Q | | D | 5 | G4(MP2)-D-v1[b] | 2.68 | 1.226 | 0.577 | 0.977 | 1.111 | | 0.776 | −0.456 | |
| Q | | D | 4 | G4(MP2)-D-v2[b] | 3.01 | 1.045 | 0.593 | 0.968 | 1.033 | | 0.796 | [0] | |
| Q[a] | | D | 5 | cc-G4(MP2)-D-v1 | 2.35 | 1.123 | 0.668 | 0.979 | 1.078 | | 0.951 | −0.487 | |
| Q[a] | | D | 4 | cc-G4(MP2)-D-v2 | 2.77 | 0.954 | 0.671 | 0.968 | 0.998 | | 0.947 | [0] | |

| | | | | | | $c_{E2,ss}$ | | $c_{E2,os}$ | $c_{CCSD-MP2}$ | | $c_T$ | $c_{Disp}$ | |
| {T,Q} | | T | 4 | G4-T-v1[c] | 1.46 | 0.787 | | | 1.064 | | 1.177 | −0.072 | |
| {T,Q} | | T | 3 | G4-T-v2[c] | 1.49 | 0.626 | | | 1.061 | | 1.103 | [0] | |
| {T,Q}[a] | | T | 4 | cc-G4-T-v1 | 0.87 | 0.707 | | | 1.053 | | 1.128 | −0.006 | |
| {T,Q}[a] | | T | 3 | cc-G4-T-v2 | 0.87 | 0.643 | | | 1.054 | | 1.126 | [0] | |
| {T,Q}[a] | | T | 3 | cc-G4-T-v2(FC) | 0.94 | 0.601 | | | 1.051 | | 1.139 | [0] | |
| {T,Q}[a] | | T | 2 | cc-G4-T-v6 | 0.90 | 0.513 | | | 1.029 | | 1.029 | [0] | |
| {T,Q}[a] | | T | 2 | cc-G4-T-v6(FC) | 0.99 | 0.604 | | | 1.017 | | 1.017 | [0] | |
| {T,Q}[a] | | T | 1 | cc-G4-T-v7 | 0.92 | 0.708 | | | 1.000 | | 1.000 | [0] | |
| {T,Q} | | T(DLPNO) | 3 | G4-T-DLPNO-v2[b] | 1.66 | 0.554 | | | 1.019 | | 1.204 | [0] | |
| {T,Q} | | Q(DLPNO) | 3 | G4-Q-DLPNO-v2[b] | 1.52 | 0.679 | | | 1.003 | | 1.185 | [0] | |
| {T,Q}[a] | | T(DLPNO) | 3 | cc-G4-T-DLPNO-v2 | 1.19 | | | | 0.999 | | 1.181 | [0] | |
| {T,Q}[a] | | T(DLPNO) | 3 | Ditto (with MP2(FC)) | 1.21 | | | | 0.999 | | 1.193 | [0] | |
| {T,Q}[a] | | Q-DLPNO | 3 | cc-G4-Q-DLPNO-v2 | 1.00 | | | | 0.998 | | 1.156 | [0] | |
| {T,Q}[a] | | Q(DLPNO) | 3 | Ditto (with MP2(FC)) | 1.00 | | | | 0.997 | | 1.173 | [0] | |

| | | | | | | $c_{E2,ss}$ | | $c_{E3}$ | $c_{CCSD-MP3}$ | | $c_T$ | $c_{Disp}$ | |
| Q | T | D | 6 | G4(MP3)-D-v1[b] | 1.65 | 1.284 | | 1.089 | 1.119 | | 0.411 | −0.185 | |
| Q | T | D | 6 | cc-G4(MP3)-D-v1 | 1.37 | 1.196 | | 1.021 | 1.121 | | 0.572 | −0.235 | |
| Q | T | | 4 | MP2.X-Q[b] | 3.28 | 1.117 | | 0.824 | [0] | | [0] | −0.062 | |
| Q[a] | T | | 4 | cc-MP2.X-Q | 2.89 | 0.879 | | 0.689 | [0] | | [0] | −0.052 | |
| {T,Q} | T | D | 3 | G4(scsMP2.X)-D-v8[b] | 1.84 | 0.633 | | 0.965 | 1.026 | | 1.026 | [0] | |
| {T,Q}[a] | T | D | 3 | cc-G4(scsMP2.X)-D-v8 | 1.44 | 1.617 | | 0.926 | 1.003 | | 1.003 | [0] | |

| | | | | | | $c_{E2}$ | | | $c_{CCSD-MP2}$ | | $c_T$ | $c_{Disp}$ | |
| MP2-F12 | | CCSD(T) | | | | | | | | | | | |
| Q | | T | 1 | cc-G4-F12-T-v2 | 1.03 | 1.000 | | | 1.000 | | 1.000 | [0] | |
| T | | T(DLPNO) | | cc-G4-F12-T-DLPNO | 1.19 | 1.000 | | | 1.000 | | 1.204 | [0] | |
| (d) | (g) | 6-31G(d) | 6 | G4 | 2.52 | | | | | | | | Y |

Table 2. continued

| MP2 | MP3 | CCSD(T) | $N_{param}$ | method | WTMAD2 | $c_{(E2,os)}$ | $c_{(E2,ss)}$ | $c_{E2,os}$ | $c_{E2,ss}$ | $c_{E(C,CCSD)}$ | $c_T$ | $c_{Disp}$ | HLC |
|---|---|---|---|---|---|---|---|---|---|---|---|---|---|
| (e) | | 6-31G(d) | 6 | G4(MP2) | 2.96 | | | | | | | | Y |
| (f) | (h) | 6-31+G(d') | 3 | CBS-QB3 | 3.10 | | | | | | | | |

[a]aug-cc-pwCV$m$Z(-PP) for both HF/CBS and $E_2$ correlation energy. [b]The results from the conventional G4(MP2)-XK[35] and standard cWFT methods were obtained from ref 38 based on the GMTKN55 geometries. [c]The WTMAD2 component breakdown is for 1320 reactions (+63 more than in our previous work[38]). [d]G3LargeXP denotes the G3Large basis set used in the G3 method but the 2df and 3d2f polarization functions are replaced by 3df and 4d2f on the second-row atoms, respectively. [e]G3MP2LargeXP is a variant of G3LargeXP in which the core polarization functions of G3LargeXP are deleted. [f]CBSB3 stands for the 6-311++G(3d2f,2df,2p) basis set, where 3d2f functions are added on second-row atoms, 2df on first-row atoms, and 2p on hydrogen. [g]MP4/6-31G(2df,p) and MP4/6-31+G(d'). [h]MP4(SDQ)/CBSB4. CBSB4 denotes the 6-31+G(d(f),p) basis set with a d function on selected second-row atoms, an f function on first- and second-row atoms, and a p function on hydrogen.

This leads to cc-G4(MP3)-D-v1 with WTMAD2 = 1.37 kcal/mol and six parameters. The overall WTMAD2 improvement relative to the standard G4(MP3)-D-v1 is 0.28 kcal/mol, when using larger basis sets and including the CV energy in cc-G4(MP3)-D-v1; it is an accumulative amelioration rising from most subsets as they slightly benefit from the larger basis sets and the CV inclusion (Table S5 in the Supporting Information). Eliminating the dispersion term raises WTMAD2 by 0.17 kcal/mol (cc-G4(MP3)-D-v2). Setting $c_T$ = $c_{CCSD-MP3}$ yields WTMAD2 = 1.65 kcal/mol (cc-G4(MP3)-D-v8). In addition, when $c_T$ = $c_{CCSD-MP3}$ = $c_{E3}$, we obtain WTMAD2 = 1.69 kcal/mol, but with just three parameters.

**Correlation-Consistent DLPNO-CCSD(T)-Based cWFT Methods.** One way to improve computational efficiency would be to substitute DLPNO-CCSD(T) for CCSD(T), leading to two-tier cWFTs that combine RI-MP2 for the larger basis sets with a smaller basis set CCSD(T)-MP2 correction computed at the DLPNO-CCSD(T) level. While the latter asymptotically scales linearly with system size, RI-MP2 still has $O(n^5)$ scaling; however, as discussed in Section 5 of Weigend et al.,[56] and demonstrated in Table 5 there, the prefactor of the $O(n^5)$ term is small enough that it does not dominate until molecules of the size of 44-alkane are reached.

These correlation-consistent cWFT approaches follow the previously reported energy expressions (Table 2 of ref 38), but substituting RI-MP2/aug-cc-pwCV{T,Q}Z(-PP) extrapolations without frozen cores. For example, cc-G4-Q-DLPNO-v1 then entails the following corrections: the difference [DLPNO-CCSD−DLPNO-MP2]/def2-QZVPP, the extrapolated RI-MP2/aug-cc-pwCV{T,Q}Z(-PP), the triples excitation DLPNO-(T)/def2-QZVPP, and the dispersion. The only common terms between cc-G4-Q-DLPNO-v1 and the standard G4-Q-DLPNO-v1 are the post-MP2 ones provided by DLPNO-CCSD(T) and DLPNO-MP2 using def2-QZVPP with the frozen-core approximation.

We obtain the lowest WTMAD2, 1.001 kcal/mol, for cc-G4-Q-DLPNO-v1; in view of the very small $c_{Disp}$ = 0.04, omitting the dispersion correction unsurprisingly leads to an essentially identical WTMAD2 = 1.005 kcal/mol for cc-G4-Q-DLPNO-v2, compared to 1.52 kcal/mol for G4-Q-DLPNO-v2. According to Table S6 in the Supporting Information, the half-dozen subsets that present the largest reduction in WTMAD2 are RG18, PCONF21, HAL59, AMINO20X4, HEAVY28, and BH76, the main difference from our observations for cc-G4-T-v2 above being the presence of PCONF21. (These latter calculations are computationally feasible for DLPNO-CCSD(T)/def2-QZVPP but proved too demanding for canonical CCSD(T) even with the smaller def2-TZVPP basis set.) Once again, the improvement for W4-11 only contributes −0.022 kcal/mol to the change in WTMAD2, but an improvement of RMSD from 2.9 to 0.9 kcal/mol is most definitely significant for small-molecule thermochemistry.

The WTMAD2 component breakdown reveals that all five top-level subsets are ameliorated (cc-G4-Q-DLPNO-v2 vs G4-Q-DLPNO-v2), with conformers (0.152 kcal/mol) and intermolecular interactions (0.214 kcal/mol) accounting together for 2/3 of the total improvement in ΔWTMAD2 of 0.52 kcal/mol. The three fitted parameters of cc-G4-Q-

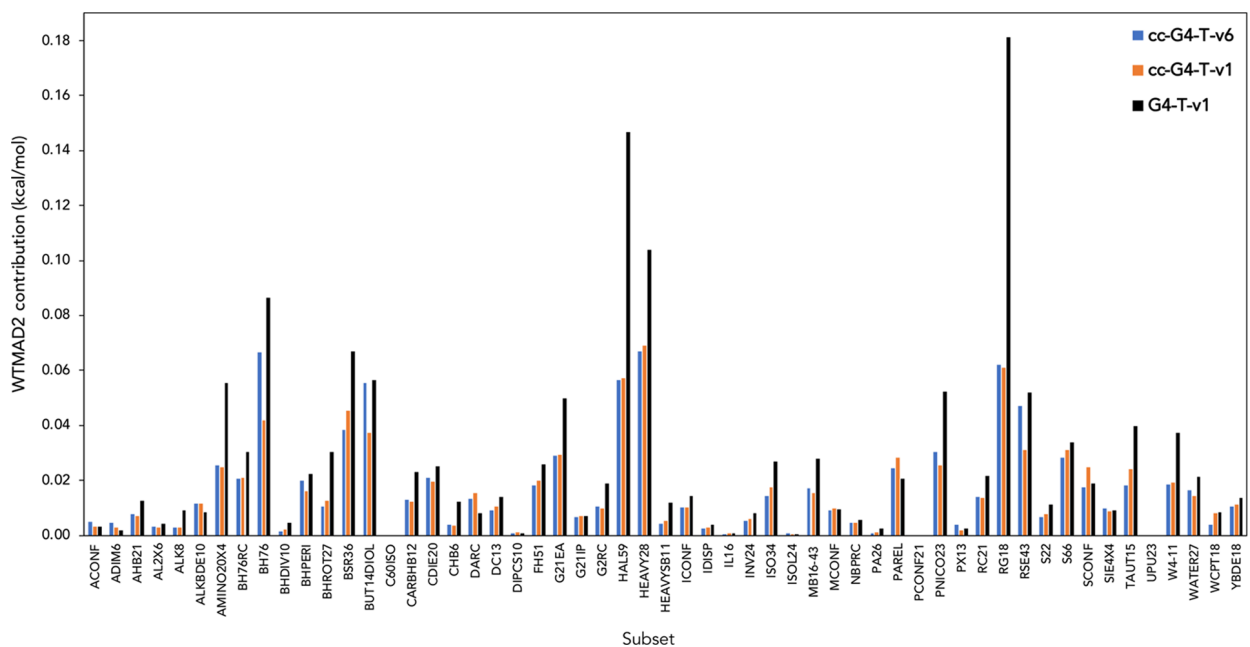

**Figure 2.** Contribution of each subset of the GMTKN55 database to the WTMAD2 (kcal/mol) for the most accurate two-tier standard (G4-T-v1, four parameters) and correlation-consistent (cc-G4-T-v1 and cc-G4-T-v6, four and two parameters, respectively) cWFT methods.

DLPNO-v2 are $c_{MP2/CBS}$ = 0.554, $c_{CCSD-MP2}$ = 0.998, and $c_T$ = 1.156; for the standard G4-Q-DLPNO-v2, we obtained $c_{MP2/CBS}$ = 0.513, $c_{CCSD-MP2}$ = 1.003, and $c_T$ = 1.185. The slight change for $c_{MP2/CBS}$ may be attributed to the larger basis sets and the inclusion of the inner-shell correlation in the $E_2$ energy. The cc-G4-T-DLPNO-v2 and cc-G4-D-DLPNO-v2 methods follow with WTMAD2 = 1.18 and 1.83 kcal/mol, respectively; when eliminating dispersion, the statistics similarly change negligibly. We note that $c_{CCSD-MP2}$ is essentially unity, while $c_T$ takes on larger values; this compensates for the known fact[117,118] that (T) in DLPNO-CCSD(T) is known to not fully recover the triples because of the neglect of off-diagonal Fock matrix elements.

This latter effect can be gauged by substituting the more elaborate (and resource-hungry) DLPNO-CCSD($T_1$) for DLPNO-CCSD(T). We considered the cc-G4-D-DLPNO-$T_1$ and cc-G4-T-DLPNO-$T_1$ variants based on DLPNO-CCSD-($T_1$)/def2-SVPD and DLPNO-CCSD($T_1$)/def2-TZVPP, respectively. The best overall result we obtained was WTMAD2 = 1.11 kcal/mol for cc-G4-T-DLPNO-$T_1$-v1, which marginally increases by 0.03 kcal/mol when discarding the dispersion correction (cc-G4-T-DLPNO-$T_1$-v2). DLPNO-CCSD($T_1$) is much more demanding in resources—particularly I/O bandwidth—than DLPNO-CCSD(T), and as in our previous study,[38] we find (somewhat surprisingly) that it offers no significant accuracy advantage in the present context: For the same number of systems, cc-G4-T-DLPNO-$T_1$-v2 yields WTMAD2 = 1.14 kcal/mol, compared to 1.13 kcal/mol for cc-G4-T-DLPNO-v2. The individual contributions to the WTMAD2 per subset are summarized in Table S7.

In the DLPNO-CCSD(T)-based variants, we had applied the RIJCOSX approximation to avoid a scenario for large molecules where the SCF step would dominate overall CPU time. Thus far, we had only applied the default GridXS2 grid in RIJCOSX. Did this fairly coarse grid introduce significant error? To elucidate this point, we repeated the DLPNO-CCSD(T)/def2-TZVPPD and DLPNO-MP2/def2-TZVPPD calculations using GRIDX9 (the most stringent built-in option) for the RIJCOSX step. By way of illustration, for phenol and for melatonin, this added just 7.0 and 3.8%, respectively, to the wall clock time for DLPNO-CCSD(T)/def2-TZVPP. The new terms ($E_{C,DLPNO-CCSD} - E_{C,DLPNO-MP2}$) and $E_{C,DLPNO-(T)}$ were then substituted in cc-G4-T-DLPNO-v2; the resulting WTMAD2 = 1.140 kcal/mol is 0.053 kcal/mol lower than for the default GRIDXS2, narrowing the gap with the canonical CCSD(T)-based cc-G4-T to 0.17 kcal/mol. The difference is concentrated in the conformer subsets ACONF, AMINO20X4, and PCONF21 (together −0.037 kcal/mol), as well as HEAVY28; the rest of the subsets are not notably affected. The optimized parameters for cc-G4-T-DLPNO-v2 with GRIDX9 marginally change to $c_{MP2/CBS}$ = 0.61717, $c_{CCSD-MP2}$ = 1.00053, $c_T$ = 1.1754, while for cc-G4-T-DLPNO-v2 with default grid, those are $c_{MP2/CBS}$ = 0.61192, $c_{CCSD-MP2}$ = 0.99895, $c_T$ = 1.18091. A detailed comparison of the WTMAD2 breakdown per subset for cc-G4-T-DLPNO with GRIDX9 vs default GRIDXS2 is reported in the Supporting Information (see Table S8). Clearly, one can set $c_{CCSD-MP2}$ = 1.0 with impunity, reducing the number of empirical parameters to just two. In light of the very small cost penalty, we recommend using GRIDX9 throughout if one avails oneself of RIJCOSX.

Regarding the explicitly correlated RI-MP2-F12-based approximations, we begin by considering additivity approximations of the following form

$$E = E_{HF(CABS)/cc-pVQZ-F12} + c_{E2}E_{RI-MP2-F12/cc-pVQZ-F12} \\ + c_{CCSD-MP2}E_{CCSD-MP2/def2-TZVP} \\ + c_T E_{C,(T)/def2-TZVP} \quad (8)$$

The cc-G4-F12-T-v2 with $c_{E2} = c_{CCSD-MP2} = c_T = 1.0$ reaches WTMAD2 = 1.030 kcal/mol and a comparison with cc-G4(FC)-T-v2 ($E_2$/aug-cc-pwCV{T,Q}(-PP) with Schwenke-style extrapolation) shows that thermochemistry and intermolecular reactions are not affected, though some deterioration is seen for barrier heights and reactions involving large molecules (see Table 2). A detailed comparison by subsets can be found in Table S9 of the SI.

Reducing the RI-MP2-F12 basis set to TZ only marginally increases WTMAD2 by 0.014 kcal/mol [cc-G4-F12-T-v1]. This is a testament to the ability of explicitly correlated methods[119−122] to drastically speed up basis set convergence, typically by two or three angular momentum steps over their orbital counterparts.[123] We attempted RI-MP2-F12/cc-pV-{T,Q}Z-F12 extrapolation using the Schwenke coefficient $c_{E2/CBS}$ = 0.446336 from Table 6 of Hill et al.[92] {$E_2$/CBS = $E_2(L + 1) + c_{E2/CBS}[E_2(L + 1) − E_2(L)]$}, but found that WTMAD2 slightly increases to 1.048 kcal/mol. Likely, seeing an advantage to larger basis sets in the F12 step would require tightening the post-MP2 steps as well. Thus, TZ quality basis sets in RI-MP2-F12 and the post-MP2 terms represent a "sweet spot" for cc-G4-F12-T-v1.

Next, we address the question whether DLPNO-CCSD(T) can replace the canonical CCSD(T) terms in the cc-G4-F12 variants. Substituting $E_{[DLPNO-CCSD–DLPNO-MP2]/def2-TZVPP}$ and $E_{DLPNO-CCSD(T),T/def2-TZVPP}$ in eq 8 and setting $c_{E2}$ = $c_{DLPNO-CCSD–DLPNO-MP2}$ = $c_{T,DLPNO-CCSD(T)}$ = 1, WTMAD2 increases to 1.44 kcal/mol. However, if we optimize all three parameters, WTMAD2 drops to 1.189 kcal/mol for $c_{E2}$ = 0.994; $c_{DLPNO-CCSD–DLPNO-MP2}$ = 0.993; $c_{T,DLPNO-CCSD(T)}$ = 1.237. That is, while the first two parameters can be set to unity, (T) clearly needs to be scaled up as we saw above. Doing so leads to the single-parameter methods cc-G4-F12-T-DLPNO-v2 with WTMAD2 = 1.194 kcal/mol for $c_{T,DLPNO-CCSD(T)}$ = 1.204 and $c_{E2}$ = $c_{DLPNO-CCSD–DLPNO-MP2}$ = 1.0. This is analogous to cc-G4-T-DLPNO-v2 (WTMAD2 = 1.193 kcal/mol) with its triples coefficient being equal to 1.181; in both cases, this reflects[118] that (T) in DLPNO-CCSD(T)—which should actually have been called DLPNO-CCSD($T_0$) and is referred to as such in ref 118—does not recoup the full thermochemical contribution of triples owing to the $T_0$ (neglect of off-diagonal Fock matrix elements) approximation.

**Final Selected Methods.** The hierarchy of the cc-G4-type cWFT closely parallels that of the standard G4-type cWFT methods, especially for the two-tier approaches. When a CCSD(T)/def2-TZVP or DLPNO-CCSD(T)/def2-TZVPP component is present in these cc-cWFTs, WTMAD2 values below 1 kcal/mol can be reached for GMTKN55.

The top performers are correlation-consistent two-tier methods. These include [cc-G4-T-v6] cc-G4-T with WTMAD2 = 0.90 kcal/mol and only two parameters, followed by the low-cost DLPNO-CCSD(T)-based [cc-G4-Q-DLPNO-v2] cc-G4-Q-DLPNO with WTMAD2 equal to 1.00 kcal/mol, and finally, [cc-G4-F12-T-v1] cc-G4-F12-T WTMAD2 = 1.04 kcal/mol. To put this in perspective, the top performers among the def2-based G4-type methods[38] likewise belonged to the two-tier family, though G4(MP2)-XK-T was found to have WTMAD2 values 0.09 and 0.08 kcal/mol lower than G4-Q-DLPNO and G4-T, respectively. These energy differences are marginal, below 0.1 kcal/mol, and in our previous work, we considered 1257 reactions for the CCSD(T)/def2-TZVP-based methods. Even when extending to 1320 reactions (i.e., completing CCSD(T)/def2-TZVP for some large species that we were unable to finish in the previous study), these trends are not affected (Table 2).

Combining the $E_2$/aug-cc-pwCVQZ(-PP) with an MP3/def2-TZVPP component and a low-cost CCSD(T)/def2-SVPD step yields [cc-G4(MP3)-D-v1] cc-G4(MP3)-D with WTMAD2 = 1.37 kcal/mol and six parameters. This result is competitive with the CCSD(T)/def2-TZVP- and MP2/def2-QZVPP-based two-tier G4-type methods. An efficient RI-MP3 algorithm would dramatically reduce the computational cost of G4(MP3)-D, as it would eliminate the I/O overhead of the MP3/def2-TZVPP step ($O(N_{el}^6)$ scaling over system size).

By reducing the basis set for CCSD(T) to the split-valence def2-SVSP, we arrive at the lowest-cost two-tier methods, cc-G4-D-DLPNO (cc-G4-D-DLPNO-v2) with WTMAD2 = 1.84 kcal/mol and cc-G4(MP2)-XK-D (cc-G4(MP2)-XK-D-v1) with WTMAD2 = 2.21 kcal/mol. When omitting the CCSD(T) step entirely, we obtain as best overall approximations cc-MP2.X-Q (without dispersion) with WTMAD2 = 2.89 kcal/mol and cc-MP2.X-T (with dispersion) with WTMAD2 = 3.09 kcal/mol. We summarize the suggested correlation-consistent cWFT methods in Table 3, and in Figure 3, we depict the overall performance of selected G4-type and cc-G4-type cWFTs, standard cWFTs, and double-hybrid DFTs.

Table 3. Summary of Recommended Correlation-Consistent cWFT Methods

| methods | WTMAD2 (kcal/mol) | parameters |
| --- | --- | --- |
| cc-G4-T | 0.90 | 2 |
| cc-G4-Q-DLPNO | 1.00 | 3;2[a] |
| cc-G4-F12-T | 1.04 | |
| cc-G4(MP2)-XK-T | 1.18$_5$ | 6 |
| cc-G4-T-DLPNO | 1.19$_3$ | 3;2[a] |
| cc-G4-F12-T-DLPNO | 1.19$_4$ | 1 |
| cc-G4(MP3)-D | 1.37 | 6 |
| cc-G4-D-DLPNO | 1.84 | 3 |
| cc-G4(MP2)-XK-D | 2.21 | 7 |

[a]If one fixes $c_{CCSD-MP2}$ = 1; effect on WTMAD2 invisible to the precision given.

The cc-G4-type methods provide a low-cost and quite accurate approximation to the CCSD(T) electronic energy at the complete basis set limit. As such, an inherent limitation is the shortcomings of the CCSD(T) method itself for species with strong static correlation; post-CCSD(T) approaches are currently out of reach for the larger species in GMTKN55. This may limit the applicability of these cWFT methods to transition metals, although they may still be valuable for applications on second- and third-row precious-metal catalysts (see, e.g., refs 124, 125 for reviews), where static correlation effects are milder.[126,127] This issue will be investigated in future work. Suffice to say for now that for main-group systems such as those in the GMTKN55 dataset, correlation-consistent cWFT approaches—both ccCA and the parametrized approaches offered here—represent felicitous combinations of moderate cost and fairly high accuracy, with WTMAD2 values less than half what can be achieved by the best empirical double hybrids.[43,44,106]

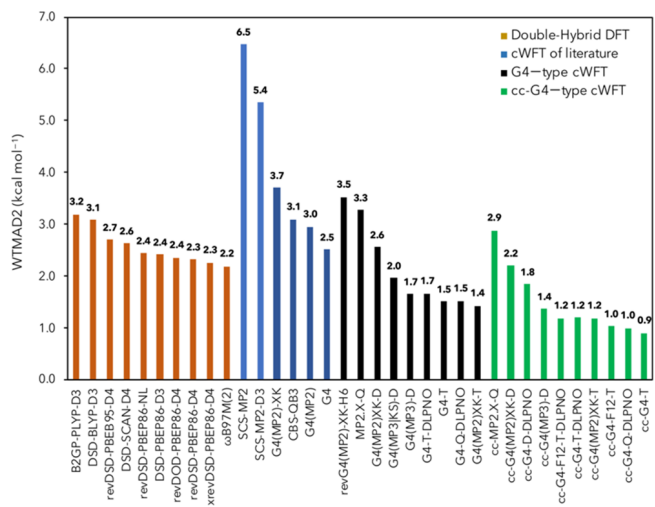

Figure 3. Overall performance of selected composite methods and double-hybrid DFT over the GMTKN55 database based on the weighted mean absolute deviation (WTMAD2 in kcal/mol).

Table 4. Wall Clock Times (Min) of the Top-Performing Composite Wavefunction Methods on Two 18-Core Intel Xeon Gold 6240 CPUs (2.30 GHz) for Species from the W4-11 and ADIM6 Datasets

| species | cc-G4-T | cc-G4-T-DLPNO | cc-G4-D-DLPNO | cc-G4-F12-T-DLPNO | W4 |
|---|---|---|---|---|---|
| HCl | 2.4 | 2.9 | 2.3 | 1.3 | 5.9 |
| HS | 4.2 | 5.8 | 4.2 | 3.1 | 7.0 |
| $H_2O$ | 2.3 | 3.2 | 2.4 | 1.7 | 18.1 |
| $ALH_3$ | 7.4 | 8.2 | 7.3 | 1.9 | 23.9 |
| $BH_3$ | 3.3 | 3.9 | 3.3 | 1.5 | 42.4 |
| $PH_3$ | 7.1 | 8.9 | 7.3 | 3.0 | 65.7 |
| HCN | 6.9 | 9.1 | 7.2 | 3.7 | 82.0 |
| HOF | 6.8 | 11.6 | 8.4 | 6.1 | 862.2 |
| ethane...ethane | 11.4 | 13.5 | 11.6 | 5.4 | |
| propane...propane | 39.9 | 43.4 | 38.6 | 14.2 | |
| butane...butane | 109.3 | 115.3 | 106.5 | 29.5 | |
| pentane...pentane | 255.7 | 197.4 | 179.4 | 56.7 | |
| hexane...hexane | 582.7 | 334.9 | 306.6 | 95.1 | |
| heptane...heptane | 832.5 | 518.3 | 475.7 | 151.4 | |

**Timing Comparisons of Selected Methods.** At the request of a reviewer, we now briefly compare the computational cost of the top-performing cWFT methods with each other and with the high-accuracy W4 approach. To keep the playing field level, identical hardware is used for each calculation in these comparisons. Some timing data are presented in Table 4. The post-CCSD(T) steps in W4—particularly CCSDT(Q), which scales as $O(n^4N^5)$ with system size, and CCSDTQ, which scales as $O(n^4N^6)$—quickly become the dominant factor in the W4 CPU time, and it is hence not surprising that cc-G4-T and cc-G4-T-DLPNO are 1−2 orders of magnitude faster. (Their costliest steps are CCSD(T)/def2-TZVP and DLPNO-CCSD(T)/def2-TZVPPD, respectively.)

For such small molecules (e.g., those in the W4-11 test set), cc-G4-T-DLPNO can actually be more expensive than the canonical counterpart, cc-G4-T. However, for larger molecules that are no longer amenable to W4 calculations with present-day hardware, cc-G4-T-DLPNO does offer an increasing speedup over its canonical sibling as molecules grow larger (illustrated in Table 4 for the n-alkane dimer series), owing to the nearly linear CPU time scaling of DLPNO-CCSD(T). Additionally, replacing RI-MP2 by RI-MP2-F12 offers a very substantial further speedup. For example, for n-heptane dimer, the wall clock time is just 2.5 h for cc-G4-F12-T-DLPNO, compared to 8.6 h cc-G4-T-DLPNO and 13.9 h for cc-G4-T. This overall improvement is primarily attributed to the accelerated basis set convergence of RI-MP2-F12 in cc-G4-F12-T-DLPNO, allowing us to stop at cc-pVTZ-F12 for that step. Execution times for other W4-11 species are listed in the Supporting Information.

**Recommendations Concerning ZPE, Thermal, and Relativistic Corrections.** All GMTKN55 reference values we considered here and in ref 38 are free of relativistic corrections and at the bottom of the well.

For "turnkey" computational thermochemistry, G4, G4-(MP2), G4(MP2)-XK, and the like include automatic computation of the zero-point vibration energy (ZPVE) and thermal corrections, typically at the same level as used for the geometry optimization. All frequencies are then scaled by a uniform scale factor appropriate for the zero-point vibrational energy: see refs 128−130 for detailed discussions. Suffice to say that, as first pointed out in refs 131, 132, scaling factors for matching experimental infrared and Raman spectra are too small for zero-point vibrational energies, which require less "scaling down" as the anharmonic corrections in them are proportionally only 25% as large. As shown in ref 128, CCSD(T) near the basis set limit has a scaling factor of essentially 1.0 for harmonic frequencies, but about 0.987 for ZPVEs: that such a rudimentary approach as uniform scaling works at all is due to anharmonicity constants being roughly proportional to the corresponding harmonic frequencies.[128]

In G4 and G4(MP2), the level of theory chosen is B3LYP/6-31G(2df,p), frequencies scaled by 0.9854: in ref 128, this level was found to lead to an RMSD of 0.10 kcal/mol on the ZPVE part of the HFREQ2014 database. G4(MP2)-XK employs the BMK functional[133] instead (frequencies scaled by 0.9766 for ZPVE, 0.9647 for entropy, and 0.9791 for enthalpy corrections): BMK by design[133] will be more reliable for locating transition states but leaves something to be desired in terms of harmonic frequencies.

One might instead consider the most recent range-separated hybrid functionals ωB97X-V[110] or ωB97M-V,[109] which for many properties are best-in-class.[43] Another alternative, however, would be to employ double-hybrid functionals, for which multiple codes have analytical first- and even second-derivative implementations. For codes that lack RI-MP2, double hybrids come at a premium; the most widely used such code, Gaussian 16, in any case cannot evaluate ωB97X-V or ωB97M-V. If RI-MP2 is available, however, double hybrids may actually be faster than range-separated hybrids, except for quite large molecules. The original DSD-PBEP86 functional[107,134] was shown[128] to be capable of reproducing the HFREQ2014 harmonic frequencies database[128] to an RMSD of just 10 cm$^{-1}$, which translates into a ZPVE contribution of 10/(2 × 349.76) = 0.014 kcal/mol per mode, the scaling factor for harmonic frequencies being different from 1.0 only in name; for zero-point vibrational energies, a scaling factor of 0.9834 yields an RMSD of just 0.05 kcal/mol! The revised revDSD-PBEP86-D3(BJ) functional[44] is superior to the original across the board for energetic properties: for harmonic frequencies and zero-point vibrational energies, it performs similarly to the original. For the sake of completeness, we evaluated the ZPVE scaling factors for revDSD-PBEP86-D3(BJ) with the exact same procedure as in ref 128, for three commonly used def2 basis sets: we found 0.9841 for def2-TZVP, 0.9822 for def2-TZVPP, and 0.9827 for def2-QZVP, with RMSD values of 0.048, 0.046, and 0.044 kcal/mol, respectively. Clearly, revDSD-PBEP86-D3(BJ)/def2-TZVP is a felicitous compromise between accuracy and computational cost, and we recommend it as our geometry optimization and harmonic frequencies level of theory.

For spin−orbit coupling, like for G4, G4(MP2), and G4(MP2)XK, we recommend using the atomic first-order splitting corrections commonly applied, which are obtained from experimental fine structures (e.g., the Harvard atomic spectra database[135]): compilations can be found in the Supporting Information of the present paper as well as of ref 136.

This leaves the scalar relativistic contribution. For applications like noncovalent interactions and conformer equilibria of biomolecules, this can safely be omitted, while for heavy-element compounds, one may need to use relativistic Hamiltonians from the ground up. In between these two regimes, we know that at least for some of the W4-11 atomization energies (e.g., $SO_3$, $SiF_4$), scalar relativistic corrections can reach the 1 kcal/mol regime.[72] (We recently proposed[137] a very simple additive model to rationalize and semiquantitatively predict the magnitudes of relativistic corrections in terms of changes in s population.)

One fairly inexpensive workaround would be to apply the same relativistic correction as ccCA, namely, $E$[DKH2-MP2/cc-pVTZ-DK] − $E$[MP2/cc-pVTZ], where cc-pVTZ-DK is a recontraction of the cc-pVTZ basis set for the second-order Douglas−Kroll−Hess (DKH2)[138] Hamiltonian. We will benchmark relativistic correction schemes in greater detail in the near future, in the context of studies on transition-metal compounds and catalysts.

## CONCLUSIONS

We have extended the hierarchy of the composite wave-function methods by (a) considering inner-shell correlation in the second-order Møller−Plesset step and (b) replacing the Karlsruhe basis sets by augmented correlation-consistent core−valence basis sets of triple and quadruple $\zeta$ quality. The resulting cc-G4-type methods reach WTMAD2 statistics below 1 kcal/mol for the large and diverse GMTKN55 benchmark suite.

Somewhat to our surprise, the lion's share of the improvement did not come from core−core and core−valence contributions, but from enhancements in the basis sets. Nevertheless, the extra cost of including the additional electrons in the RI-MP2 step is such a small component of the overall CPU time that there is no downside to including them.

A thorough investigation of each subset's contribution showed that the statistical improvement for the two-tier methods lies in the larger molecules, e.g., improved energies of amino acid conformers, barrier heights of pericyclic reactions, binding energies in halogenated dimers, relative energies in the tri- and tetrapeptide conformers, binding energies of noncovalently bound dimers, and relative energies in tautomers. In contrast, amelioration in the three-tier methods is seen across the board and is not concentrated in specific subsets.

The minimally empirical cc-G4-T breaches the 1 kcal/mol threshold, with WTMAD2 = 0.90 kcal/mol and only two fitted parameters for the chemically diverse GMTKN55 database. The two-tier cc-G4-T-v7 (inspired by ccCA) reaches WTMAD2 of 0.92 kcal/mol; it is available for the spd block of H-Rn in the PTE, and $c_{MP2/CBS}$ is the single parameter since $c_{CCSD-MP2} = c_T = 1.0$. As post-HF corrections, the extrapolated $E_2$/aug-cc-pwCV{T,Q}Z(-PP) and the CCSD(T)/def2-TZVP components are used. (A putative nonempirical variant of cc-G4-T-v7 in which we replace the lone remaining empirical parameter by $c_{MP2/CBS} = ((4/3)^3 − 1)^{-1} = 0.730$ would be quite similar to the nonrelativistic part of a ccCA calculation, except for including MP2 core−valence correlation throughout: it yields WTMAD2 = 0.93 kcal/mol.)

The corresponding efficient DLPNO-CCSD(T)-based cWFTs are also very attractive owing to the replacement of the costly CCSD(T) by the linear-scaling DLPNO-CCSD(T) component. The lower-cost cc-G4-T-DLPNO reaches a WTMAD2 of 1.20 kcal/mol at a moderate computational cost: by way of illustration, a melatonin conformer takes 24.2 h of wall clock time on two 8-core Intel Haswell processors (Xeon CPU E5-2630v3 clocked at 2.40 GHz).

The three-tier cc-G4(MP3)-D method is in the same accuracy range as cc-G4-T-DLPNO. Said method has an MP2/aug-cc-pwCVQZ(-PP) term, an MP3/def2-TZVPP component, and the lower-cost CCSD(T)/def2-SVPD; this combination yields a WTMAD2 of 1.37 kcal/mol. An efficient RI-MP3 algorithm will render cc-G4(MP3)-D more efficient and dramatically reduce its cost and I/O overhead.

Finally, cc-G4-F12-T-DLPNO, which combines explicitly correlated MP2-F12/cc-pVTZ-F12 with DLPNO-CCSD(T)/def2-TZVPP and has just one empirical parameter, is an accurate and fairly inexpensive alternative for compounds where cc-pVTZ-F12 or cc-pVTZ-PP-F12 basis sets are available for all elements.

All in all, the cc-G4-type approaches offer a material improvement in terms of accuracy over G4-T and similar cWFT approaches, at a comparable computational cost.

## ASSOCIATED CONTENT

### Supporting Information

The Supporting Information is available free of charge at https://pubs.acs.org/doi/10.1021/acs.jctc.0c01106.

> Sample input files and postprocessing scripts for top performers: cc-G4-T, cc-G4-Q-DLPNO, cc-G4-F12-T-DLPNO, cc-G4(MP2)-XK-T, cc-G4-T-DLPNO, cc-G4-(MP3)-D, cc-G4-D-DLPNO, and cc-G4(MP2)-XK-D (ZIP)

## AUTHOR INFORMATION

**Corresponding Author**

Jan M. L. Martin − *Department of Organic Chemistry, Weizmann Institute of Science, 7610001 Rehovot, Israel;* orcid.org/0000-0002-0005-5074; Email: gershom@weizmann.ac.il

**Author**

Emmanouil Semidalas − *Department of Organic Chemistry, Weizmann Institute of Science, 7610001 Rehovot, Israel;* orcid.org/0000-0002-4464-4057

Complete contact information is available at:
https://pubs.acs.org/10.1021/acs.jctc.0c01106

**Funding**

This research was supported by the Israel Science Foundation (grants 1358/15 and 1969/20) and by the Weizmann Institute's SABRA (Supporting Advanced Basic Research) program; the latter was supported by a grant from the Estate of Emile Mimran. The work of E.S. on this scientific paper was supported by the Onassis Foundation—Scholarship ID: F ZP 052-1/2019-2020.

**Notes**

The authors declare no competing financial interest.

## ■ ACKNOWLEDGMENTS

The authors thank Golokesh Santra and Dr. Mark A. Iron for helpful discussions, and Dr. Mark Vilensky (scientific computing manager of CHEMFARM) for technical assistance.

## ■ REFERENCES

(1) Curtiss, L. A.; Raghavachari, K.; Redfern, P. C.; Rassolov, V.; Pople, J. A. Gaussian-3 (G3) Theory for Molecules Containing First and Second-Row Atoms. *J. Chem. Phys.* **1998**, *109*, 7764−7776.

(2) Baboul, A. G.; Curtiss, L. A.; Redfern, P. C.; Raghavachari, K. Gaussian-3 Theory Using Density Functional Geometries and Zero-Point Energies. *J. Chem. Phys.* **1999**, *110*, 7650−7657.

(3) Curtiss, L. A.; Redfern, P. C.; Raghavachari, K. Gaussian-4 Theory. *J. Chem. Phys.* **2007**, *126*, No. 084108.

(4) Curtiss, L. A.; Redfern, P. C.; Raghavachari, K. Gaussian-4 Theory Using Reduced Order Perturbation Theory. *J. Chem. Phys.* **2007**, *127*, No. 124105.

(5) Curtiss, L. A.; Raghavachari, K.; Trucks, G. W.; Pople, J. A. Gaussian-2 Theory for Molecular Energies of First- and Second-row Compounds. *J. Chem. Phys.* **1991**, *94*, 7221−7230.

(6) Pople, J. A.; Head-Gordon, M.; Fox, D. J.; Raghavachari, K.; Curtiss, L. A. Gaussian-1 Theory: A General Procedure for Prediction of Molecular Energies. *J. Chem. Phys.* **1989**, *90*, 5622−5629.

(7) Curtiss, L. A.; McGrath, M. P.; Blaudeau, J.; Davis, N. E.; Binning, R. C.; Radom, L. Extension of Gaussian-2 Theory to Molecules Containing Third-row Atoms Ga−Kr. *J. Chem. Phys.* **1995**, *103*, 6104−6113.

(8) Montgomery, J. A.; Frisch, M. J.; Ochterski, J. W.; Petersson, G. A. A Complete Basis Set Model Chemistry. VI. Use of Density Functional Geometries and Frequencies. *J. Chem. Phys.* **1999**, *110*, 2822−2827.

(9) Montgomery, J. A.; Frisch, M. J.; Ochterski, J. W.; Petersson, G. A. A Complete Basis Set Model Chemistry. VII. Use of the Minimum Population Localization Method. *J. Chem. Phys.* **2000**, *112*, 6532−6542.

(10) Zhao, Y.; Lynch, B. J.; Truhlar, D. G. Doubly Hybrid Meta DFT: New Multi-Coefficient Correlation and Density Functional Methods for Thermochemistry and Thermochemical Kinetics. *J. Phys. Chem. A* **2004**, *108*, 4786−4791.

(11) Zhao, Y.; Lynch, B. J.; Truhlar, D. G. Multi-Coefficient Extrapolated Density Functional Theory for Thermochemistry and Thermochemical Kinetics. *Phys. Chem. Chem. Phys.* **2005**, *7*, 43.

(12) Zhao, Y.; Xia, L.; Liao, X.; He, Q.; Zhao, M. X.; Truhlar, D. G. Extrapolation of High-Order Correlation Energies: The WMS Model. *Phys. Chem. Chem. Phys.* **2018**, *20*, 27375−27384.

(13) DeYonker, N. J.; Cundari, T. R.; Wilson, A. K. The Correlation Consistent Composite Approach (CcCA): An Alternative to the Gaussian-n Methods. *J. Chem. Phys.* **2006**, *124*, No. 114104.

(14) DeYonker, N. J.; Wilson, B. R.; Pierpont, A. W.; Cundari, T. R.; Wilson, A. K. Towards the Intrinsic Error of the Correlation Consistent Composite Approach (CcCA). *Mol. Phys.* **2009**, *107*, 1107−1121.

(15) Peterson, C.; Penchoff, D. A.; Wilson, A. K. Prediction of Thermochemical Properties Across the Periodic Table. In *Annual Reports in Computational Chemistry*; Elsevier, 2016; Vol. *12*, pp 3−45.

(16) Martin, J. M. L.; de Oliveira, G. Towards Standard Methods for Benchmark Quality Ab Initio Thermochemistry—W1 and W2 Theory. *J. Chem. Phys.* **1999**, *111*, 1843−1856.

(17) Parthiban, S.; Martin, J. M. L. Assessment of W1 and W2 Theories for the Computation of Electron Affinities, Ionization Potentials, Heats of Formation, and Proton Affinities. *J. Chem. Phys.* **2001**, *114*, 6014−6029.

(18) Martin, J. M. L.; Parthiban, S. W1 and W2 Theories, and Their Variants: Thermochemistry in the KJ/Mol Accuracy Range. In *Quantum-Mechanical Prediction of Thermochemical Data*; Cioslowski, J., Ed.; Kluwer Academic Publishers: Dordrecht, 2002; Vol. *22*, pp 31−65.

(19) Karton, A.; Rabinovich, E.; Martin, J. M. L.; Ruscic, B. W4 Theory for Computational Thermochemistry: In Pursuit of Confident Sub-KJ/Mol Predictions. *J. Chem. Phys.* **2006**, *125*, No. 144108.

(20) Karton, A.; Martin, J. M. L. Explicitly Correlated Wn Theory: W1-F12 and W2-F12. *J. Chem. Phys.* **2012**, *136*, No. 124114.

(21) Sylvetsky, N.; Peterson, K. A.; Karton, A.; Martin, J. M. L. Toward a W4-F12 Approach: Can Explicitly Correlated and Orbital-Based Ab Initio CCSD(T) Limits Be Reconciled? *J. Chem. Phys.* **2016**, *144*, No. 214101.

(22) Chan, B.; Radom, L. W1X-1 and W1X-2: W1-Quality Accuracy with an Order of Magnitude Reduction in Computational Cost. *J. Chem. Theory Comput.* **2012**, *8*, 4259−4269.

(23) Chan, B.; Radom, L. W2X and W3X-L: Cost-Effective Approximations to W2 and W4 with kJ Mol$^{-1}$ Accuracy. *J. Chem. Theory Comput.* **2015**, *11*, 2109−2119.

(24) Tajti, A.; Szalay, P. G.; Császár, A. G.; Kállay, M.; Gauss, J.; Valeev, E. F.; Flowers, B. A.; Vázquez, J.; Stanton, J. F. HEAT: High Accuracy Extrapolated Ab Initio Thermochemistry. *J. Chem. Phys.* **2004**, *121*, 11599−11613.

(25) Bomble, Y. J.; Vázquez, J.; Kállay, M.; Michauk, C.; Szalay, P. G.; Császár, A. G.; Gauss, J.; Stanton, J. F. High-Accuracy Extrapolated Ab Initio Thermochemistry. II. Minor Improvements to the Protocol and a Vital Simplification. *J. Chem. Phys.* **2006**, *125*, No. 064108.

(26) Harding, M. E.; Vázquez, J.; Ruscic, B.; Wilson, A. K.; Gauss, J.; Stanton, J. F. High-Accuracy Extrapolated Ab Initio Thermochemistry. III. Additional Improvements and Overview. *J. Chem. Phys.* **2008**, *128*, No. 114111.

(27) Feller, D.; Peterson, K. A.; Ruscic, B. Improved Accuracy Benchmarks of Small Molecules Using Correlation Consistent Basis Sets. *Theor. Chem. Acc.* **2013**, *133*, 1407.

(28) Dixon, D. A.; Feller, D.; Peterson, K. A. A Practical Guide to Reliable First Principles Computational Thermochemistry Predictions Across the Periodic Table. In *Annual Reports in Computational Chemistry*; Elsevier, 2012; Vol. *8*, pp 1−28.

(29) Feller, D.; Peterson, K. A.; Dixon, D. A. The Impact of Larger Basis Sets and Explicitly Correlated Coupled Cluster Theory on the Feller−Peterson−Dixon Composite Method. *Annu. Rep. Comput. Chem.* **2016**, *12*, 47−48.

(30) Raghavachari, K.; Trucks, G. W.; Pople, J. A.; Head-Gordon, M. A Fifth-Order Perturbation Comparison of Electron Correlation Theories. *Chem. Phys. Lett.* **1989**, *157*, 479−483.

(31) Watts, J. D.; Gauss, J.; Bartlett, R. J. Coupled-Cluster Methods with Noniterative Triple Excitations for Restricted Open-Shell Hartree−Fock and Other General Single Determinant Reference Functions. Energies and Analytical Gradients. *J. Chem. Phys.* **1993**, *98*, 8718−8733.

(32) Patel, P.; Wilson, A. K. Domain-Based Local Pair Natural Orbital Methods within the Correlation Consistent Composite Approach. *J. Comput. Chem.* **2020**, *41*, 800−813.

(33) Curtiss, L. A.; Raghavachari, K.; Redfern, P. C.; Pople, J. A. Assessment of Gaussian-2 and Density Functional Theories for the Computation of Enthalpies of Formation. *J. Chem. Phys.* **1997**, *106*, 1063−1079.

(34) Curtiss, L. A.; Redfern, P. C.; Raghavachari, K.; Pople, J. A. Assessment of Gaussian-2 and Density Functional Theories for the Computation of Ionization Potentials and Electron Affinities. *J. Chem. Phys.* **1998**, *109*, 42−55.

(35) Chan, B.; Karton, A.; Raghavachari, K. G4(MP2)-XK: A Variant of the G4(MP2)-6X Composite Method with Expanded Applicability for Main-Group Elements up to Radon. *J. Chem. Theory Comput.* **2019**, *15*, 4478−4484.

(36) Weigend, F.; Ahlrichs, R. Balanced Basis Sets of Split Valence, Triple Zeta Valence and Quadruple Zeta Valence Quality for H to Rn: Design and Assessment of Accuracy. *Phys. Chem. Chem. Phys.* **2005**, *7*, 3297−3305.

(37) Goerigk, L.; Hansen, A.; Bauer, C.; Ehrlich, S.; Najibi, A.; Grimme, S. A Look at the Density Functional Theory Zoo with the Advanced GMTKN55 Database for General Main Group Thermochemistry, Kinetics and Noncovalent Interactions. *Phys. Chem. Chem. Phys.* **2017**, *19*, 32184−32215.

(38) Semidalas, E.; Martin, J. M. L. Canonical and DLPNO-Based G4(MP2)XK-Inspired Composite Wave Function Methods Parametrized against Large and Chemically Diverse Training Sets: Are They More Accurate and/or Robust than Double-Hybrid DFT? *J. Chem. Theory Comput.* **2020**, *16*, 4238−4255.

(39) Grimme, S. Semiempirical Hybrid Density Functional with Perturbative Second-Order Correlation. *J. Chem. Phys.* **2006**, *124*, No. 034108.

(40) Sancho-García, J. C.; Adamo, C. Double-Hybrid Density Functionals: Merging Wavefunction and Density Approaches to Get the Best of Both Worlds. *Phys. Chem. Chem. Phys.* **2013**, *15*, 14581.

(41) Su, N. Q.; Xu, X. The XYG3 Type of Doubly Hybrid Density Functionals. *Wiley Interdiscip. Rev.: Comput. Mol. Sci.* **2016**, *6*, 721−747.

(42) Brémond, E.; Ciofini, I.; Sancho-García, J. C.; Adamo, C. Nonempirical Double-Hybrid Functionals: An Effective Tool for Chemists. *Acc. Chem. Res.* **2016**, *49*, 1503−1513.

(43) Martin, J. M. L.; Santra, G. Empirical Double-Hybrid Density Functional Theory: A 'Third Way' in Between WFT and DFT. *Isr. J. Chem.* **2020**, *60*, 787−804.

(44) Santra, G.; Sylvetsky, N.; Martin, J. M. L. Minimally Empirical Double-Hybrid Functionals Trained against the GMTKN55 Database: RevDSD-PBEP86-D4, RevDOD-PBE-D4, and DOD-SCAN-D4. *J. Phys. Chem. A* **2019**, *123*, 5129−5143.

(45) Peterson, K. A.; Dunning, T. H. Accurate Correlation Consistent Basis Sets for Molecular Core−Valence Correlation Effects: The Second Row Atoms Al−Ar, and the First Row Atoms B−Ne Revisited. *J. Chem. Phys.* **2002**, *117*, 10548−10560.

(46) Sylvetsky, N.; Martin, J. M. L. Probing the Basis Set Limit for Thermochemical Contributions of Inner-Shell Correlation: Balance of Core-Core and Core-Valence Contributions. *Mol. Phys.* **2019**, *117*, 1078−1087.

(47) Martin, J. M. L.; Sundermann, A.; Fast, P. L.; Truhlar, D. G. Thermochemical Analysis of Core Correlation and Scalar Relativistic Effects on Molecular Atomization Energies. *J. Chem. Phys.* **2000**, *113*, 1348−1358.

(48) Řezáč, J.; Hobza, P. Describing Noncovalent Interactions beyond the Common Approximations: How Accurate Is the "Gold Standard," CCSD(T) at the Complete Basis Set Limit? *J. Chem. Theory Comput.* **2013**, *9*, 2151−2155.

(49) Kesharwani, M. K.; Manna, D.; Sylvetsky, N.; Martin, J. M. L. The X40×10 Halogen Bonding Benchmark Revisited: Surprising Importance of (n −1)d Subvalence Correlation. *J. Phys. Chem. A* **2018**, *122*, 2184−2197.

(50) Hollett, J. W.; Gill, P. M. W. The Two Faces of Static Correlation. *J. Chem. Phys.* **2011**, *134*, No. 114111.

(51) Karton, A.; Rabinovich, E.; Martin, J. M. L.; Ruscic, B. W4 Theory for Computational Thermochemistry: In Pursuit of Confident Sub-KJ/Mol Predictions. *J. Chem. Phys.* **2006**, *125*, No. 144108.

(52) Karton, A.; Taylor, P. R.; Martin, J. M. L. Basis Set Convergence of Post-CCSD Contributions to Molecular Atomization Energies. *J. Chem. Phys.* **2007**, *127*, No. 064104.

(53) Thorpe, J. H.; Lopez, C. A.; Nguyen, T. L.; Baraban, J. H.; Bross, D. H.; Ruscic, B.; Stanton, J. F. High-Accuracy Extrapolated Ab Initio Thermochemistry. IV. A Modified Recipe for Computational Efficiency. *J. Chem. Phys.* **2019**, *150*, No. 224102.

(54) Frisch, M. J.; Trucks, G. W.; Schlegel, H. B.; Scuseria, G. E.; Robb, M. A.; Cheeseman, J. R.; Scalmani, G.; Barone, V.; Petersson, G. A.; Nakatsuji, H.; Li, X.; Caricato, M.; Marenich, A. V.; Bloino, J.; Janesko, B. G.; Gomperts, R.; Mennucci, B.; Hratchian, H. P.; Ortiz, J. V.; Izmaylov, A. F.; Sonnenberg, J. L.; Williams-Young, D.; Ding, F.; Lipparini, F.; Egidi, F.; Goings, J.; Peng, B.; Petrone, A.; Henderson, T.; Ranasinghe, D.; Zakrzewski, V. G.; Gao, J.; Rega, N.; Zheng, G.; Liang, W.; Hada, M.; Ehara, M.; Toyota, K.; Fukuda, R.; Hasegawa, J.; Ishida, M.; Nakajima, T.; Honda, Y.; Kitao, O.; Nakai, H.; Vreven, T.; Throssell, K.; Montgomery, J. A. J.; Peralta, J. E.; Ogliaro, F.; Bearpark, M. J.; Heyd, J. J.; Brothers, E. N.; Kudin, K. N.; Staroverov, V. N.; Keith, T. A.; Kobayashi, R.; Normand, J.; Raghavachari, K.; Rendell, A. P.; Burant, J. C.; Iyengar, S. S.; Tomasi, J. C.; Cossi, M.; Millam, J. M.; Klene, M.; Adamo, C.; Cammi, R.; Ochterski, J. W.; Martin, R. L.; Morokuma, K.; Farkas, O.; Foresman, J. B.; Fox, D. J. Gaussian 16, revision C.01.; Gaussian, Inc.: Wallingford, CT, 2016.

(55) Häser, M. Møller−Plesset (MP2) Perturbation Theory for Large Molecules. *Theor. Chim. Acta* **1993**, *87*, 147−173.

(56) Weigend, F.; Häser, M.; Patzelt, H.; Ahlrichs, R. RI-MP2: Optimized Auxiliary Basis Sets and Demonstration of Efficiency. *Chem. Phys. Lett.* **1998**, *294*, 143−152.

(57) Shao, Y.; Gan, Z.; Epifanovsky, E.; Gilbert, A. T. B.; Wormit, M.; Kussmann, J.; Lange, A. W.; Behn, A.; Deng, J.; Feng, X.; Ghosh, D.; Goldey, M.; Horn, P. R.; Jacobson, L. D.; Kaliman, I.; Khaliullin, R. Z.; Kuś, T.; Landau, A.; Liu, J.; Proynov, E. I.; Rhee, Y. M.; Richard, R. M.; Rohrdanz, M. A.; Steele, R. P.; Sundstrom, E. J.; Woodcock, H. L.; Zimmerman, P. M.; Zuev, D.; Albrecht, B.; Alguire, E.; Austin, B.; Beran, G. J. O.; Bernard, Ya.; Berquist, E.; Brandhorst, K.; Bravaya, K. B.; Brown, S. T.; Casanova, D.; Chang, C.-M.; Chen, Y.; Chien, S. H.; Closser, K. D.; Crittenden, D. L.; Diedenhofen, M.; DiStasio, R. A.; Do, H.; Dutoi, A. D.; Edgar, R. G.; Fatehi, S.; Fusti-Molnar, L.; Ghysels, A.; Golubeva-Zadorozhnaya, A.; Gomes, J.; Hanson-Heine, M. W. D.; Harbach, P. H. P.; Hauser, A. W.; Hohenstein, E. G.; Holden, Z. C.; Jagau, T.-C.; Ji, H.; Kaduk, B.; Khistyaev, K.; Kim, J.; Kim, J.; King, R. A.; Klunzinger, P.; Kosenkov, D.; Kowalczyk, T.; Krauter, C. M.; Lao, K. U.; Laurent, A. D.; Lawler, K. V.; Levchenko, S. V.; Lin, C. Y.; Liu, F.; Livshits, E.; Lochan, R. C.; Luenser, A.; Manohar, P.; Manzer, S. F.; Mao, S.-P.; Mardirossian, N.; Marenich, A. V.; Maurer, Sa.; Mayhall, N. J.; Neuscamman, E.; Oana, C. M.; Olivares-Amaya, R.; O'Neill, D. P.; Parkhill, Ja.; Perrine, T. M.; Peverati, R.; Prociuk, A.; Rehn, D. R.; Rosta, E.; Russ, N. J.; Sharada, S. M.; Sharma, S.; Small, D. W.; Sodt, A.; Stein, T.; Stück, D.; Su, Y.-C.; Thom, A. J. W.; Tsuchimochi, T.; Vanovschi, V.; Vogt, L.; Vydrov, O.; Wang, T.; Watson, Ma.; Wenzel, J.; White, A.; Williams, C. F.; Yang, J.; Yeganeh, S.; Yost, S. R.; You, Z.-Q.; Zhang, I. Y.; Zhang, X.; Zhao, Y.; Brooks, B. R.; Chan, G. K. L.; Chipman, D. M.; Cramer, C. J.; Goddard, W. A.; Gordon, M. S.; Hehre, W. J.; Klamt, A.; Schaefer, H. F.; Schmidt, M. W.; Sherrill, C. D.; Truhlar, D. G.; Warshel, A.; Xu, X.; Aspuru-Guzik, A.; Baer, R.; Bell, A. T.; Besley, N.; Chai, J.-D.; Dreuw, A.; Dunietz, B. D.; Furlani, T. R.; Gwaltney, S. R.; Hsu, C.-P.; Jung, Y.; Kong, J.; Lambrecht, D. S.; Liang, W.; Ochsenfeld, C.; Rassolov, V. A.; Slipchenko, L. V.; Subotnik, J. E.; Van Voorhis, T.; Herbert, J. M.; Krylov, A. I.; Gill, P. M. W.; Head-Gordon, M. Advances in Molecular Quantum Chemistry Contained in the Q-Chem 4 Program Package. *Mol. Phys.* **2015**, *113*, 184−215.

(58) Pinski, P.; Riplinger, C.; Valeev, E. F.; Neese, F. Sparse Maps—A Systematic Infrastructure for Reduced-Scaling Electronic Structure Methods. I. An Efficient and Simple Linear Scaling Local MP2

Method That Uses an Intermediate Basis of Pair Natural Orbitals. *J. Chem. Phys.* **2015**, *143*, No. 034108.

(59) Riplinger, C.; Sandhoefer, B.; Hansen, A.; Neese, F. Natural Triple Excitations in Local Coupled Cluster Calculations with Pair Natural Orbitals. *J. Chem. Phys.* **2013**, *139*, No. 134101.

(60) Guo, Y.; Riplinger, C.; Becker, U.; Liakos, D. G.; Minenkov, Y.; Cavallo, L.; Neese, F. Communication: An Improved Linear Scaling Perturbative Triples Correction for the Domain Based Local Pair-Natural Orbital Based Singles and Doubles Coupled Cluster Method [DLPNO-CCSD(T)]. *J. Chem. Phys.* **2018**, *148*, No. 011101.

(61) Neese, F.; Wennmohs, F.; Becker, U.; Riplinger, C. The ORCA Quantum Chemistry Program Package. *J. Chem. Phys.* **2020**, *152*, No. 224108.

(62) Werner, H.-J.; Knowles, P. J.; Manby, F. R.; Black, J. A.; Doll, K.; Heßelmann, A.; Kats, D.; Köhn, A.; Korona, T.; Kreplin, D. A.; Ma, Q.; Miller, T. F.; Mitrushchenkov, A.; Peterson, K. A.; Polyak, I.; Rauhut, G.; Sibaev, M. The Molpro Quantum Chemistry Package. *J. Chem. Phys.* **2020**, *152*, No. 144107.

(63) Dunning, T. H. Gaussian Basis Sets for Use in Correlated Molecular Calculations. I. The Atoms Boron through Neon and Hydrogen. *J. Chem. Phys.* **1989**, *90*, 1007−1023.

(64) Kendall, R. A.; Dunning, T. H.; Harrison, R. J. Electron Affinities of the First-Row Atoms Revisited. Systematic Basis Sets and Wave Functions. *J. Chem. Phys.* **1992**, *96*, 6796−6806.

(65) Weigend, F.; Köhn, A.; Hättig, C. Efficient Use of the Correlation Consistent Basis Sets in Resolution of the Identity MP2 Calculations. *J. Chem. Phys.* **2002**, *116*, 3175−3183.

(66) Yousaf, K. E.; Peterson, K. A. Optimized Complementary Auxiliary Basis Sets for Explicitly Correlated Methods: Aug-Cc-PVnZ Orbital Basis Sets. *Chem. Phys. Lett.* **2009**, *476*, 303−307.

(67) Hättig, C. Optimization of Auxiliary Basis Sets for RI-MP2 and RI-CC2 Calculations: Core−Valence and Quintuple-ζ Basis Sets for H to Ar and QZVPP Basis Sets for Li to Kr. *Phys. Chem. Chem. Phys.* **2005**, *7*, 59−66.

(68) Peterson, K. A.; Yousaf, K. E. Molecular Core-Valence Correlation Effects Involving the Post-d Elements Ga−Rn: Benchmarks and New Pseudopotential-Based Correlation Consistent Basis Sets. *J. Chem. Phys.* **2010**, *133*, No. 174116.

(69) Peterson, K. A.; Figgen, D.; Goll, E.; Stoll, H.; Dolg, M. Systematically Convergent Basis Sets with Relativistic Pseudopotentials. II. Small-Core Pseudopotentials and Correlation Consistent Basis Sets for the Post-*d* Group 16−18 Elements. *J. Chem. Phys.* **2003**, *119*, 11113−11123.

(70) Metz, B.; Stoll, H.; Dolg, M. Small-Core Multiconfiguration-Dirac−Hartree−Fock-Adjusted Pseudopotentials for Post-d Main Group Elements: Application to PbH and PbO. *J. Chem. Phys.* **2000**, *113*, 2563−2569.

(71) Hill, J. G.; Peterson, K. A. Correlation Consistent Basis Sets for Explicitly Correlated Wavefunctions: Pseudopotential-Based Basis Sets for the Post-d Main Group Elements Ga−Rn. *J. Chem. Phys.* **2014**, *141*, No. 094106.

(72) Karton, A.; Daon, S.; Martin, J. M. L. W4-11: A High-Confidence Benchmark Dataset for Computational Thermochemistry Derived from First-Principles W4 Data. *Chem. Phys. Lett.* **2011**, *510*, 165−178.

(73) Dolg, M.; Cao, X. Relativistic Pseudopotentials: Their Development and Scope of Applications. *Chem. Rev.* **2012**, *112*, 403−480.

(74) Pritchard, B. P.; Altarawy, D.; Didier, B. T.; Gibson, T. D.; Windus, T. L. A New Basis Set Exchange: An Open, Up-to-Date Resource for the Molecular Sciences Community. *J. Chem. Inf. Model.* **2019**, *59*, 4814−4820.

(75) Hill, J. G. *Correlation Consistent Basis Set Repository*. http://www.grant-hill.group.shef.ac.uk/ccrepo/ (accessed November 3, 2020).

(76) Chan, B.; Karton, A.; Raghavachari, K.; Radom, L. Restricted-Open-Shell G4(MP2)-Type Procedures. *J. Phys. Chem. A* **2016**, *120*, 9299−9304.

(77) Grimme, S.; Antony, J.; Ehrlich, S.; Krieg, H. A Consistent and Accurate Ab Initio Parametrization of Density Functional Dispersion Correction (DFT-D) for the 94 Elements H-Pu. *J. Chem. Phys.* **2010**, *132*, No. 154104.

(78) Grimme, S.; Ehrlich, S.; Goerigk, L. Effect of the Damping Function in Dispersion Corrected Density Functional Theory. *J. Comput. Chem.* **2011**, *32*, 1456−1465.

(79) Liakos, D. G.; Sparta, M.; Kesharwani, M. K.; Martin, J. M. L.; Neese, F. Exploring the Accuracy Limits of Local Pair Natural Orbital Coupled-Cluster Theory. *J. Chem. Theory Comput.* **2015**, *11*, 1525−1539.

(80) Dolg, M. Relativistic Effective Core Potentials. In *Handbook of Relativistic Quantum Chemistry*; Liu, W., Ed.; Springer: Berlin, Heidelberg, 2017; pp 449−478.

(81) Neese, F.; Wennmohs, F.; Hansen, A.; Becker, U. Efficient, Approximate and Parallel Hartree−Fock and Hybrid DFT Calculations. A 'Chain-of-Spheres' Algorithm for the Hartree−Fock Exchange. *Chem. Phys.* **2009**, *356*, 98−109.

(82) Izsa′k, R.; Neese, F. An Overlap Fitted Chain of Spheres Exchange Method. *J. Chem. Phys.* **2011**, *135*, No. 144105.

(83) Krack, M.; Köster, A. M. An Adaptive Numerical Integrator for Molecular Integrals. *J. Chem. Phys.* **1998**, *108*, 3226−3234.

(84) Weigend, F. Accurate Coulomb-Fitting Basis Sets for H to Rn. *Phys. Chem. Chem. Phys.* **2006**, *8*, 1057−1065.

(85) Hellweg, A.; Hättig, C.; Höfener, S.; Klopper, W. Optimized Accurate Auxiliary Basis Sets for RI-MP2 and RI-CC2 Calculations for the Atoms Rb to Rn. *Theor. Chem. Acc.* **2007**, *117*, 587−597.

(86) Rappoport, D.; Furche, F. Property-Optimized Gaussian Basis Sets for Molecular Response Calculations. *J. Chem. Phys.* **2010**, *133*, No. 134105.

(87) Womack, J. C.; Manby, F. R. Density Fitting for Three-Electron Integrals in Explicitly Correlated Electronic Structure Theory. *J. Chem. Phys.* **2014**, *140*, No. 044118.

(88) Peterson, K. A.; Adler, T. B.; Werner, H.-J. Systematically Convergent Basis Sets for Explicitly Correlated Wavefunctions: The Atoms H, He, B-Ne, and Al-Ar. *J. Chem. Phys.* **2008**, *128*, No. 084102.

(89) Weigend, F. A Fully Direct RI-HF Algorithm: Implementation, Optimised Auxiliary Basis Sets, Demonstration of Accuracy and Efficiency. *Phys. Chem. Chem. Phys.* **2002**, *4*, 4285−4291.

(90) Yousaf, K. E.; Peterson, K. A. Optimized Auxiliary Basis Sets for Explicitly Correlated Methods. *J. Chem. Phys.* **2008**, *129*, No. 184108.

(91) Knizia, G.; Adler, T. B.; Werner, H.-J. Simplified CCSD(T)-F12 Methods: Theory and Benchmarks. *J. Chem. Phys.* **2009**, *130*, No. 054104.

(92) Hill, J. G.; Peterson, K. A.; Knizia, G.; Werner, H.-J. Extrapolating MP2 and CCSD Explicitly Correlated Correlation Energies to the Complete Basis Set Limit with First and Second Row Correlation Consistent Basis Sets. *J. Chem. Phys.* **2009**, *131*, No. 194105.

(93) Morgante, P.; Peverati, R. ACCDB: A Collection of Chemistry Databases for Broad Computational Purposes. *J. Comput. Chem.* **2019**, *40*, 839−848.

(94) Sure, R.; Hansen, A.; Schwerdtfeger, P.; Grimme, S. Comprehensive Theoretical Study of All 1812 $C_{60}$ Isomers. *Phys. Chem. Chem. Phys.* **2017**, *19*, 14296−14305.

(95) Kruse, H.; Mladek, A.; Gkionis, K.; Hansen, A.; Grimme, S.; Sponer, J. Quantum Chemical Benchmark Study on 46 RNA Backbone Families Using a Dinucleotide Unit. *J. Chem. Theory Comput.* **2015**, *11*, 4972−4991.

(96) Powell, M. J. D. *The BOBYQA Algorithm for Bound Constrained Optimization without Derivatives*, Cambridge NA Report NA2009/06; University of Cambridge: Cambridge, U.K., 2009.

(97) Huber, P. J.; Ronchetti, E. M. *Robust Statistics*; Wiley Series in Probability and Statistics; John Wiley & Sons, Inc: Hoboken, NJ, 2009.

(98) Kállay, M.; Nagy, P. R.; Mester, D.; Rolik, Z.; Samu, G.; Csontos, J.; Csóka, J.; Szabó, P. B.; Gyevi-Nagy, L.; Hégely, B.; Ladjánszki, I.; Szegedy, L.; Ladóczki, B.; Petrov, K.; Farkas, M.; Mezei, P. D.; Ganyecz, Á. The MRCC Program System: Accurate Quantum

Chemistry from Water to Proteins. *J. Chem. Phys.* **2020**, *152*, No. 074107.

(99) Kállay, M.; Surján, P. R. Higher Excitations in Coupled-Cluster Theory. *J. Chem. Phys.* **2001**, *115*, 2945−2954.

(100) Kállay, M.; Gauss, J. Approximate Treatment of Higher Excitations in Coupled-Cluster Theory. *J. Chem. Phys.* **2005**, *123*, No. 214105.

(101) Halkier, A.; Helgaker, T.; Jørgensen, P.; Klopper, W.; Olsen, J. Basis-Set Convergence of the Energy in Molecular Hartree-Fock Calculations. *Chem. Phys. Lett.* **1999**, *302*, 437−446.

(102) Schwenke, D. W. The Extrapolation of One-Electron Basis Sets in Electronic Structure Calculations: How It Should Work and How It Can Be Made to Work. *J. Chem. Phys.* **2005**, *122*, No. 014107.

(103) Martin, J. M. L. A Simple "Range Extender" for Basis Set Extrapolation Methods for MP2 and Coupled Cluster Correlation Energies. *AIP Conf. Proc.* **2018**, *2040*, No. 020008.

(104) Grimme, S. Improved Third-Order Møller−Plesset Perturbation Theory. *J. Comput. Chem.* **2003**, *24*, 1529−1537.

(105) Grimme, S. Improved Second-Order Møller−Plesset Perturbation Theory by Separate Scaling of Parallel- and Antiparallel-Spin Pair Correlation Energies. *J. Chem. Phys.* **2003**, *118*, 9095−9102.

(106) Mardirossian, N.; Head-Gordon, M. Survival of the Most Transferable at the Top of Jacob's Ladder: Defining and Testing the ωB97M(2) Double Hybrid Density Functional. *J. Chem. Phys.* **2018**, *148*, No. 241736.

(107) Kozuch, S.; Martin, J. M. L. Spin-Component-Scaled Double Hybrids: An Extensive Search for the Best Fifth-Rung Functionals Blending DFT and Perturbation Theory. *J. Comput. Chem.* **2013**, *34*, 2327−2344.

(108) Karton, A.; Tarnopolsky, A.; Lamère, J.-F.; Schatz, G. C.; Martin, J. M. L. Highly Accurate First-Principles Benchmark Data Sets for the Parametrization and Validation of Density Functional and Other Approximate Methods. Derivation of a Robust, Generally Applicable, Double-Hybrid Functional for Thermochemistry and Thermochemical. *J. Phys. Chem. A* **2008**, *112*, 12868−12886.

(109) Mardirossian, N.; Head-Gordon, M. ωB97M-V: A Combinatorially Optimized, Range-Separated Hybrid, Meta-GGA Density Functional with VV10 Nonlocal Correlation. *J. Chem. Phys.* **2016**, *144*, No. 214110.

(110) Mardirossian, N.; Head-Gordon, M. ωB97X-V: A 10-Parameter, Range-Separated Hybrid, Generalized Gradient Approximation Density Functional with Nonlocal Correlation, Designed by a Survival-of-the-Fittest Strategy. *Phys. Chem. Chem. Phys.* **2014**, *16*, 9904−9924.

(111) Zhao, Y.; Truhlar, D. G. The M06 Suite of Density Functionals for Main Group Thermochemistry, Thermochemical Kinetics, Noncovalent Interactions, Excited States, and Transition Elements: Two New Functionals and Systematic Testing of Four M06-Class Functionals and 12 Other Function. *Theor. Chem. Acc.* **2008**, *120*, 215−241.

(112) Mehta, N.; Casanova-Páez, M.; Goerigk, L. Semi-Empirical or Non-Empirical Double-Hybrid Density Functionals: Which Are More Robust? *Phys. Chem. Chem. Phys.* **2018**, *20*, 23175−23194.

(113) Burns, L. A.; Marshall, M. S.; Sherrill, C. D. Comparing Counterpoise-Corrected, Uncorrected, and Averaged Binding Energies for Benchmarking Noncovalent Interactions. *J. Chem. Theory Comput.* **2014**, *10*, 49−57.

(114) Brauer, B.; Kesharwani, M. K.; Martin, J. M. L. Some Observations on Counterpoise Corrections for Explicitly Correlated Calculations on Noncovalent Interactions. *J. Chem. Theory Comput.* **2014**, *10*, 3791−3799.

(115) Brauer, B.; Kesharwani, M. K.; Kozuch, S.; Martin, J. M. L. The S66x8 Benchmark for Noncovalent Interactions Revisited: Explicitly Correlated Ab Initio Methods and Density Functional Theory. *Phys. Chem. Chem. Phys.* **2016**, *18*, 20905−20925.

(116) Korth, M.; Grimme, S. Mindless DFT Benchmarking. *J. Chem. Theory Comput.* **2009**, *5*, 993−1003.

(117) Guo, Y.; Riplinger, C.; Becker, U.; Liakos, D. G.; Minenkov, Y.; Cavallo, L.; Neese, F. Communication: An Improved Linear Scaling Perturbative Triples Correction for the Domain Based Local Pair-Natural Orbital Based Singles and Doubles Coupled Cluster Method [DLPNO-CCSD(T)]. *J. Chem. Phys.* **2018**, *148*, No. 011101.

(118) Guo, Y.; Riplinger, C.; Liakos, D. G.; Becker, U.; Saitow, M.; Neese, F. Linear Scaling Perturbative Triples Correction Approximations for Open-Shell Domain-Based Local Pair Natural Orbital Coupled Cluster Singles and Doubles Theory [DLPNO-CCSD($T_0$/T)]. *J. Chem. Phys.* **2020**, *152*, No. 024116.

(119) Ten-no, S. Explicitly Correlated Wave Functions: Summary and Perspective. *Theor. Chem. Acc.* **2012**, *131*, 1070.

(120) Ten-no, S.; Noga, J. Explicitly Correlated Electronic Structure Theory from R12/F12 Ansätze. *Wiley Interdiscip. Rev.: Comput. Mol. Sci.* **2012**, *2*, 114−125.

(121) Kong, L.; Bischoff, F. A.; Valeev, E. F. Explicitly Correlated R12/F12 Methods for Electronic Structure. *Chem. Rev.* **2012**, *112*, 75−107.

(122) Hättig, C.; Klopper, W.; Köhn, A.; Tew, D. P. Explicitly Correlated Electrons in Molecules. *Chem. Rev.* **2012**, *112*, 4−74.

(123) Kesharwani, M. K.; Sylvetsky, N.; Köhn, A.; Tew, D. P.; Martin, J. M. L. Do CCSD and Approximate CCSD-F12 Variants Converge to the Same Basis Set Limits? The Case of Atomization Energies. *J. Chem. Phys.* **2018**, *149*, No. 154109.

(124) Gunanathan, C.; Milstein, D. Bond Activation and Catalysis by Ruthenium Pincer Complexes. *Chem. Rev.* **2014**, *114*, 12024−12087.

(125) Van Der Boom, M. E.; Milstein, D. Cyclometalated Phosphine-Based Pincer Complexes: Mechanistic Insight in Catalysis, Coordination, and Bond Activation. *Chem. Rev.* **2003**, *103*, 1759−1792.

(126) Iron, M. A.; Janes, T. Evaluating Transition Metal Barrier Heights with the Latest Density Functional Theory Exchange−Correlation Functionals: The MOBH35 Benchmark Database. *J. Phys. Chem. A* **2019**, *123*, 3761−3781.

(127) Iron, M. A.; Janes, T. Correction to "Evaluating Transition Metal Barrier Heights with the Latest Density Functional Theory Exchange−Correlation Functionals: The MOBH35 Benchmark Database. *J. Phys. Chem. A* **2019**, *123*, 6379−6380.

(128) Kesharwani, M. K.; Brauer, B.; Martin, J. M. L. Frequency and Zero-Point Vibrational Energy Scale Factors for Double-Hybrid Density Functionals (and Other Selected Methods): Can Anharmonic Force Fields Be Avoided? *J. Phys. Chem. A* **2015**, *119*, 1701−1714.

(129) Alecu, I.; Zheng, J.; Zhao, Y.; Truhlar, D. G. Computational Thermochemistry: Scale Factor Databases and Scale Factors for Vibrational Frequencies Obtained from Electronic Model Chemistries. *J. Chem. Theory Comput.* **2010**, *6*, 2872−2887.

(130) Merrick, J. P.; Moran, D.; Radom, L. An Evaluation of Harmonic Vibrational Frequency Scale Factors. *J. Phys. Chem. A* **2007**, *111*, 11683−11700.

(131) Grev, R. S.; Janssen, C. L.; Schaefer, H. F. Concerning Zero-point Vibrational Energy Corrections to Electronic Energies. *J. Chem. Phys.* **1991**, *95*, 5128−5132.

(132) Pople, J. A.; Scott, A. P.; Wong, M. W.; Radom, L. Scaling Factors for Obtaining Fundamental Vibrational Frequencies and Zero-Point Energies from HF/6-31G* and MP2/6-31G* Harmonic Frequencies. *Isr. J. Chem.* **1993**, *33*, 345−350.

(133) Boese, A. D.; Martin, J. M. L. Development of Density Functionals for Thermochemical Kinetics. *J. Chem. Phys.* **2004**, *121*, 3405−3416.

(134) Kozuch, S.; Martin, J. M. L. DSD-PBEP86: In Search of the Best Double-Hybrid DFT with Spin-Component Scaled MP2 and Dispersion Corrections. *Phys. Chem. Chem. Phys.* **2011**, *13*, 20104.

(135) Kurucz, R. L.; Bell, B. *Atomic Line Data (Kurucz CD-ROM No. 23)*; Harvard-Smithsonian Astrophysical Observatory: Cambridge, MA, 1995.

(136) Martin, J. M. L.; Sundermann, A. Correlation Consistent Valence Basis Sets for Use with the Stuttgart−Dresden−Bonn

Relativistic Effective Core Potentials: The Atoms Ga−Kr and In−Xe. *J. Chem. Phys.* **2001**, *114*, 3408−3420.

(137) Martin, J. M. L.; Sylvetsky, N. A Simple Model for Scalar Relativistic Corrections to Molecular Total Atomisation Energies. *Mol. Phys.* **2019**, *117*, 2225−2232.

(138) Reiher, M. Relativistic Douglas-Kroll-Hess Theory. *Wiley Interdiscip. Rev.: Comput. Mol. Sci.* **2012**, *2*, 139−149.